\documentclass[manuscript]{emulateapj}
\usepackage{natbib}
\usepackage{lineno}
\usepackage{multirow}
\usepackage{booktabs}
\usepackage{epsfig}
\usepackage{color}
\usepackage{ulem}
\usepackage{amsmath}
\usepackage{soul} 
\usepackage[colorinlistoftodos]{todonotes}

\def\lesssim{\mathrel{\hbox{\rlap{\hbox{\lower3pt\hbox{$\sim$}}}\hbox{\raise1pt\hbox{$<$}}}}}
\def\gtrsim{\mathrel{\hbox{\rlap{\hbox{\lower3pt\hbox{$\sim$}}}\hbox{\raise1pt\hbox{$>$}}}}}
\newcommand{\bi}{\begin{itemize}}
\newcommand{\ei}{\end{itemize}}
\newcommand{\beq}{\begin{equation}}
\newcommand{\eeq}{\end{equation}}
\newcommand{\bea}{\begin{eqnarray}}
\newcommand{\eea}{\end{eqnarray}}
\newcommand{\bqu}{\begin{quote}}
\newcommand{\equ}{\end{quote}}
\newcommand{\bctr}{\begin{center}}
\newcommand{\ectr}{\end{center}}

\newcommand{\om}{\rm \Omega_{\rm m}}

\newcommand{\oll}{\rm \Omega_\Lambda}
\newcommand{\ok}{\rm \Omega_k}

\newcommand{\white}{\color{white}}

\newcommand{\omLCDMSNonly}{0.332\pm 0.122}
\newcommand{\omLCDMSNCMB}{0.335\pm 0.042}
\newcommand{\omLCDMSNFLAT}{0.331\pm 0.038}
\newcommand{\omLCDMSNCMBBAO}{0.308\pm 0.007}

\newcommand{\olLCDMSNonly}{0.671\pm 0.163}
\newcommand{\olLCDMSNCMB}{0.670\pm 0.032}
\newcommand{\olLCDMSNFLAT}{0.669\pm 0.038}
\newcommand{\olLCDMSNCMBBAO}{0.690\pm 0.008}
\newcommand{\RLCDM}{110} 

\newcommand{\urlDR}{{\tt https://des.ncsa.illinois.edu/releases/sn}}

\newcommand{\omSNCMB}{0.321 \pm 0.018}
\newcommand{\omDESSNCMB}{0.341 \pm 0.027}    
\newcommand{\omSNCMBBAO}{0.311 \pm 0.009}
\newcommand{\omDESSNCMBBAO}{0.315 \pm 0.010} 
\newcommand{\omCMBBAO}{0.310 \pm 0.013}      

\newcommand{\wSNCMB}{-0.978 \pm 0.059}
\newcommand{\wDESSNCMB}{-0.911 \pm 0.087}    
\newcommand{\wSNCMBBAO}{-0.977 \pm 0.047}
\newcommand{\wDESSNCMBBAO}{-0.959 \pm 0.054} 
\newcommand{\wCMBBAO}{-0.988 \pm 0.072}      

\newcommand{\wdifUnblind}{0.024} 
\newcommand{\RwCDM}{81}   

\newcommand{\omwowaSNCMBBAO}{0.316 \pm 0.011}
\newcommand{\woSNCMBBAO}{-0.885 \pm 0.114}
\newcommand{\waSNCMBBAO}{-0.387 \pm 0.430}

\newcommand{\omwowaCMBBAO}{0.332 \pm 0.022}
\newcommand{\woCMBBAO}{-0.714 \pm 0.232}
\newcommand{\waCMBBAO}{-0.714 \pm 0.692}

\newcommand{\wowaFOM}{45.5}

\newcommand{\WCDM}{$w$\text{CDM}}
\newcommand{\LCDM}{$\Lambda$\text{CDM}}
\newcommand{\SAMPLENAME}{DES-SN3YR}

\newcommand{\mb}{m_B}
\newcommand{\dmuBias}{\Delta\mu_{\rm bias}}
\newcommand{\Gstep}{G_{\rm host}}
\newcommand{\Mhost}{M_{\rm host}}
\newcommand{\Msun}{M_{\odot}}
\newcommand{\vecD}{\vec{D}} 
\newcommand{\muzDATA}{\mu(z_i)_{\rm data}}
\newcommand{\muzMODEL}{\mu(z_i)_{\rm model}}
\newcommand{\sigint}{\sigma_{\rm int}}
\newcommand{\Ctot}{C_{\rm stat+syst}}

\newcommand{\sigwtot}{\sigma_{w,{\rm tot}}}
\newcommand{\sigwstat}{\sigma_{w,{\rm stat}}}
\newcommand{\sigwsyst}{\sigma_{w,{\rm syst}}}
\newcommand{\cSalt}{{\mathcal{C}}}

\newcommand{\werrStat}{0.042}
\newcommand{\werrSyst}{0.042}
\newcommand{\werrTot}{0.059}

\newcommand{\werrSystCal}{0.021}

\newcommand{\werrSystBiasAstro}{0.026}
\newcommand{\werrSystBiasSurvey}{0.023}
\newcommand{\werrSystRedshift}{0.012}
\newcommand{\werrSystNew}{0.024}

\newcommand{\NZBIN}{18}  

\newcommand{\NSPEC}{251} 
\newcommand{\NDES}{207} 
\newcommand{\NLOWZ}{122} 
\newcommand{\NTOT}{329} 
\newcommand{\NzHOST}{158} 

\newcommand{\Diff}{{\tt DiffImg}}
\newcommand{\SNANA}{{\tt SNANA}}
 
\newcommand{\BSYS}{B18}
\newcommand{\spec}{spectroscopic}
\newcommand{\mutrue}{\mu_{\rm true}}

\shorttitle{DES: Cosmological results with spectroscopically confirmed type Ia supernovae}

\begin{document}

{
\begin{nolinenumbers}
\vspace*{-\headsep}\vspace*{\headheight}
\footnotesize \hfill FERMILAB-PUB-18-590-AE \\
\vspace*{-\headsep}\vspace*{\headheight}
\footnotesize \hfill DES-2018-0368
\end{nolinenumbers}
}

\title{First Cosmology Results using Type Ia Supernovae from the 
    Dark Energy Survey: Constraints on Cosmological Parameters}


\def\andname{}

\author{
T.~M.~C.~Abbott\altaffilmark{1},
S.~Allam\altaffilmark{2},
P.~Andersen\altaffilmark{3,4},
C.~Angus\altaffilmark{5},
J.~Asorey\altaffilmark{6},
A.~Avelino\altaffilmark{7},
S.~Avila\altaffilmark{8},
B.~A.~Bassett\altaffilmark{9,10},
K.~Bechtol\altaffilmark{11},
G.~M.~Bernstein\altaffilmark{12}
% ...
\\ \vspace{0.2cm} (DES Collaboration) \\
}

\affil{$^{1}$ Cerro Tololo Inter-American Observatory, National Optical Astronomy Observatory, Casilla 603, La Serena, Chile}
\affil{$^{2}$ Fermi National Accelerator Laboratory, P. O. Box 500, Batavia, IL 60510, USA}
\affil{$^{3}$ School of Mathematics and Physics, University of Queensland,  Brisbane, QLD 4072, Australia}
\affil{$^{4}$ University of Copenhagen, Dark Cosmology Centre, Juliane Maries Vej 30, 2100 Copenhagen O}
\affil{$^{5}$ School of Physics and Astronomy, University of Southampton,  Southampton, SO17 1BJ, UK}
\affil{$^{6}$ Korea Astronomy and Space Science Institute, Yuseong-gu, Daejeon, 305-348, Korea}
\affil{$^{7}$ Harvard-Smithsonian Center for Astrophysics, 60 Garden St., Cambridge, MA 02138, USA}
\affil{$^{8}$ Institute of Cosmology and Gravitation, University of Portsmouth, Portsmouth, PO1 3FX, UK}
\affil{$^{9}$ African Institute for Mathematical Sciences, 6 Melrose Road, Muizenberg, 7945, South Africa}
\affil{$^{10}$ South African Astronomical Observatory, P.O.Box 9, Observatory 7935, South Africa}
\affil{$^{11}$ LSST, 933 North Cherry Avenue, Tucson, AZ 85721, USA}
\affil{$^{12}$ Department of Physics and Astronomy, University of Pennsylvania, Philadelphia, PA 19104, USA}
% ...


\begin{abstract}
We present the first cosmological parameter constraints using measurements of 
type Ia supernovae (SNe~Ia) from the Dark Energy Survey Supernova Program (DES-SN).
The analysis uses a subsample of \NDES\ spectroscopically confirmed SNe~Ia 
from the first three years of DES-SN, 
combined with a low-redshift sample of \NLOWZ\ SNe from the literature. 
Our ``{\SAMPLENAME}'' result from these \NTOT\ SNe~Ia is based
on a series of companion analyses and improvements covering 
SN~Ia discovery, spectroscopic selection, 
photometry, calibration, distance bias corrections, 
and evaluation of systematic uncertainties. 
For a flat \LCDM\ model we find a matter density  $\om = \omLCDMSNFLAT$.
For a flat \WCDM\ model, and combining our SN~Ia constraints with those from the 
cosmic microwave background (CMB), 
we find a dark energy equation of state $w= \wSNCMB$, and $\om= \omSNCMB$.
For a flat $w_0w_a$CDM model, and combining probes from SN~Ia, CMB and
baryon acoustic oscillations, we find $w_0=\woSNCMBBAO$ and  $w_a=\waSNCMBBAO$.
These results are in agreement with a cosmological constant and 
with previous constraints using SNe~Ia (Pantheon, JLA).
\end{abstract}


\section{Introduction}
\label{sect:intro}

Type Ia supernovae (SNe~Ia)
were used to discover the accelerating expansion of the universe (\citealt{Riess98}, \citealt{Perlmutter99}) and remain one of the key probes for understanding the nature of the mysterious ``dark energy.''
Over the last two decades, there have been considerable improvements in the 
calibration and size of samples at 
low redshift \citep{CFAjha, Hicken09b,CfA4,CSP}, 
intermediate redshift \citep{Holtzman08}, and 
high redshift \citep{snls06,essence2,Conley11,Rest14,Betoule14}.  
When combined with cosmic microwave background (CMB) data, these samples have 
been used to demonstrate that the dark energy equation of state,  $w$, 
is consistent with a cosmological constant ($w=-1$) with a precision of $\sigma_w = 0.04$.
The recent Pantheon analysis combines $>1000$ SNe~Ia from several surveys, 
resulting in $w = -1.026\pm 0.041$ \citep{pantheon}. 

The Dark Energy Survey Supernova program (DES-SN) is striving to find even 
greater numbers of SNe while 
reducing systematic uncertainties on the resulting cosmological parameters.
A top priority of this effort has been to accurately model each component of the 
DES-SN search and analysis, and to accurately simulate bias corrections for 
the SN~Ia distance measurements.
DES has also made improvements in instrumentation and calibration, including:
(i) detectors with higher $z$-band efficiency to improve measurements 
of rest frame supernova (SN) colors at high-redshift, and
(ii) extension of the photometric calibration precision over a wide color range by 
correcting each charged-coupled device (CCD)
and exposure for atmospheric variations and the
spectral energy distribution (SED) of the source (see Sect.~\ref{sect:ana}).  
These improvements enable DES-SN to make a state-of-the-art measurement of dark energy properties.

This Letter reports ``DES-SN3YR'' cosmological constraints from the 
spectroscopically confirmed SNe~Ia in the first three years of DES-SN 
in combination with a low-redshift SN~Ia sample from the literature.
The results presented here are the culmination of a series of companion papers, 
which contain details of the SN search and discovery \citep{Kessler15,Morganson18,autoscan};
spectroscopic follow-up  \citep{DAndrea18};
photometry \citep{Brout18-SMP};
calibration \citep{Burke17,Lasker18};
simulations \citep{Kessler18};
and technique to account for selection bias \citep{bbc}.
The cosmological analysis method and validation are detailed in 
\citet[\BSYS]{Brout18-ANA},
which  presents the full statistical and systematic uncertainty budget for these new results. 
\citet{Hinton18} test a new Bayesian Hierarchical Model for supernova cosmology.  
In this letter, we summarize these contributions and present our measurements of the equation-of-state ($w$) and matter density ($\om$).  
Data products used in this analysis are publicly available online.\footnote{\urlDR}  
In addition, \citet{Macaulay18} measure the Hubble constant ($H_0$) by applying these 
DES-SN3YR results to the inverse-distance-ladder method anchored to the 
standard ruler measured by baryon acoustic oscillations 
\citep[BAO]{BAODR12,carter18},
and related to the sound horizon measured with CMB data \citep{planck16}.

In  \S2 we discuss the datasets used in our analysis.  
In  \S3, we summarize the analysis pipeline.  
In  \S4, we present the cosmology results.  
In  \S5, we present our discussion and conclusions.

\section{Data Samples}
\label{sect:data}

The DES-SN sample for this analysis was collected over three 5-month-long seasons, 
from August 2013 to February 2016, using the Dark Energy Camera (DECam, \citealt{decam})  
at the Cerro Tololo Inter-American Observatory.
Ten 2.7~deg$^2$ fields were observed approximately once per week in the $griz$ filter bands (\citealt{dr1}). The average depth per visit was 23.5~mag in the eight ``shallow" fields, and 24.5~mag in the two ``deep" fields. Within 24 hours of each observation, search images were processed \citep{Morganson18},
new transients were discovered using a difference-imaging pipeline (\citealt{Kessler15}), 
and most of the subtraction artifacts were rejected with a machine-learning algorithm 
applied to image stamps (\citealt{autoscan}).

A subset of lightcurves was selected for spectroscopic follow-up observations (\citealt{DAndrea18}), 
resulting in \NSPEC\ spectroscopically confirmed SNe~Ia with redshifts $0.02 < z < 0.85$,
and $\NDES$ SNe~Ia that satisfy analysis requirements (\BSYS)
such as signal-to-noise and light curve sampling; 
this sample is called the DES-SN subset.
The spectroscopic program required a collaborative
effort coordinated across several observatories.
At low to intermediate redshifts, the primary follow-up instrument is the 
4-meter Anglo-Australian Telescope (AAT), 
which confirmed and measured redshifts for 31\% of our SN Ia sample 
(OzDES collaboration; \citealt{Yuan15,Childress17,marz}).
A variety of spectroscopic programs (\citealt{DAndrea18}) were carried out using the
European Southern Observatory Very Large Telescope,
Gemini, Gran Telescopio Canarias, Keck, Magellan, MMT, and
South African Large Telescope.

We supplement the DES-SN sample with a low-redshift ($z< 0.1$) sample, which we call the low-$z$ subset, comprising \NLOWZ\ SNe from the Harvard-Smithsonian Center for Astrophysics surveys \citep[CfA3, CfA4;][]{CfA3,CfA4} and the Carnegie Supernova Project \citep[CSP;][]{CSP, CSP2}. 
We only use samples with measured telescope+filter transmissions,
and thus CfA1 and CfA2 are not included.

\section{Analysis}
\label{sect:ana}

Supernova cosmology relies on measuring the luminosity distance ($d_L$) versus
redshift for many SNe~Ia and comparing this relation to the prediction of cosmological models.  
The distance modulus ($\mu$) is defined as
\beq \mu = 5\log[d_L/10{\rm pc}]. \label{eq:mumodel}\eeq
For a flat universe with cold dark matter density $\om$, 
dark energy density $\oll$, and speed of light $c$,
the luminosity distance to a source at redshift $z$ is given by
\beq d_L = (1+z)c\int_0^z \frac{dz^\prime}{H(z^\prime)}, 
\eeq
with
\beq 
   \label{eq:Hz}
   H(z)=H_0\left[{\om}(1+z)^3 + {\oll}(1+z)^{3(1+w)}\right]^{1/2}. 
\eeq 
Observationally, the distance modulus of a supernova is given by 
\begin{equation}
   \mu = \mb + \alpha x_1 - \beta \cSalt + M_0 +  \gamma\Gstep + \dmuBias ~.
   \label{eq:mu}
\end{equation}
For each SN~Ia, the set of $griz$ light curves are fit (\S\ref{sect:lightcurve})
to determine an amplitude ($x_0$, with $\mb \equiv -2.5\log(x_0)$),
light curve width ($x_1$), and color ($\cSalt$).
$\gamma$ describes the dependence on host-galaxy stellar mass 
($\Mhost$, \S\ref{sect:hostmass}), where
$\Gstep = +1/2$ if $\Mhost > 10^{10}\Msun$, and $\Gstep = -1/2$ if $\Mhost < 10^{10}\Msun$.
A correction for selection biases ($\dmuBias$)
is determined from simulations (\S\ref{sect:bias}).

All SNe~Ia are assumed to be characterized by $\alpha,\beta,\gamma$, and $M_0$.  
The first three parameters describe how the 
SN~Ia luminiosity is correlated with the light curve width ($\alpha x_1$), 
color ($\beta\cSalt$), and host-galaxy stellar mass ($\gamma\Gstep$).
$M_0$ accounts for both the absolute magnitude of SNe~Ia and the Hubble constant.
In the rest of this section we describe the main components of the
analysis pipeline that are needed to determine the distances (Eq.~\ref{eq:mu}) 
and cosmological parameters.

\subsection{Calibration}  \label{sect:calibration}

The DES sample is calibrated to the AB magnitude system (\citealt{oke}) 
using measurements of the Hubble Space Telescope (HST) 
CalSpec standard C26202 (\citealt{Bohlin14}). 
DES internally calibrated roughly 50 standard stars per CCD using a 
`Forward Global Calibration Method' \citep{Burke17,Lasker18}.
Improvements in calibration at the 0.01 mag ($1\%$) level are made using SED-dependent 
`chromatic corrections' to both the standard stars and to the DES-SN lightcurve photometry. 
The low-$z$ sample is calibrated to the AB system by cross-calibrating to the Pan-STARRS1 (PS1)
photometric catalogs (\citealt{scolnic15}).  
We also cross-calibrate DES to PS1 and find good agreement 
(see \S3.1.2, Fig~3 of \BSYS).

\subsection{Photometry}
To measure the SN~Ia flux for each observation,
we employ a scene modeling photometry (SMP)
approach (\citealt{Brout18-SMP}) based on previous efforts used in 
SDSS-II \citep{Holtzman08} and SNLS \citep{Astier13}.
SMP simultaneously forward models a variable SN
flux on top of a temporally constant host galaxy.
We test the precision by analyzing images that include artificial SNe~Ia, and find that
photometric biases are limited to $<0.3$\%.
Each CCD exposure is calibrated to the native photometric system of DECam, 
and zero points are determined from the standard star catalogs 
(\S\ref{sect:calibration}).

\subsection{Spectroscopy: Typing \& Redshifts}
Spectral classification was performed using both the SuperNova IDentification \citep[SNID]{SNID} 
and Superfit \citep{Superfit} software, as described in \citet{DAndrea18}. All $\NDES$ events
are spectroscopically classified as SNe~Ia.
Redshifts are obtained from host-galaxy spectra, where available, because their sharp spectral lines give more accurate redshifts ($\sigma_z\sim 5\times 10^{-4}$; \citealt{Yuan15}) than the broad SN~Ia spectroscopic features ($\sigma_z\sim 5\times 10^{-3}$). %
\NzHOST\ of the DES-SN events have host galaxy redshifts, and the rest have redshifts from the SN~Ia spectra.
For the low-$z$ sample, we use the published redshifts with a 250~km/s uncertainty from \cite{pantheon}.
Peculiar-velocity corrections are computed from \cite{Carrick15}.

\subsection{Light-curve fitting}
\label{sect:lightcurve}

To measure the SN parameters ($\mb,x_1,\cSalt$), the light curves were fit 
with \SNANA\footnote{\texttt{https://snana.uchicago.edu}} (\citealt{Kessler09SNANA})
using the SALT2 model (\citealt{Guy10}) 
and the training parameters from \cite{Betoule14}.

\begin{figure*}
\begin{centering}
\textbf{}\includegraphics[width=0.9\textwidth]{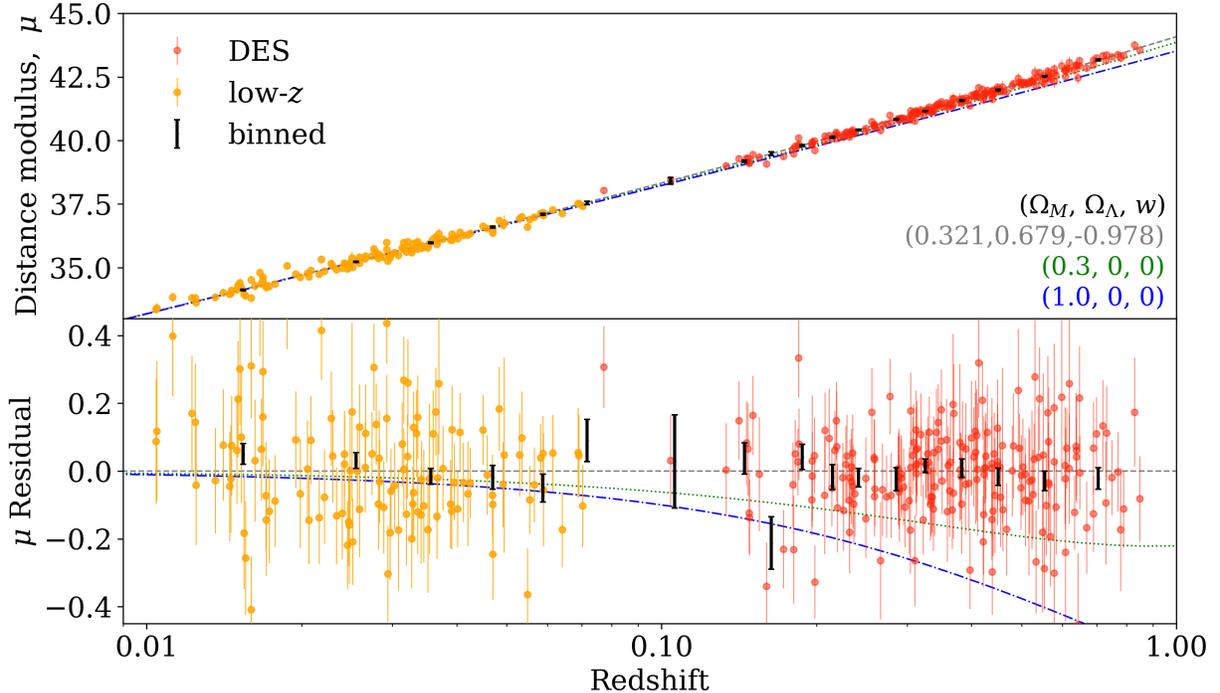}
\vspace{-0.7in}
\caption{Hubble diagram for the \SAMPLENAME\ sample. 
Top: distance modulus ($\mu$) from BBC fit 
(black bars, which are used for cosmology fits)
and for each SN (red, orange circles). The dashed gray line shows our best fit model, while the green and blue dotted lines show models with no dark energy and matter densities $\om=0.3$ and $1.0$ respectively. 
Bottom: residuals to the best fit model; $1\sigma$ error bars show 68\% confidence.}
\label{fig:hubble}
\end{centering}
\end{figure*}

\subsection{Host Galaxy Stellar Masses}
\label{sect:hostmass}
For the $\gamma\Gstep$ term in Eq.~\ref{eq:mu}, 
we first identify the host galaxy
using catalogs from Science Verification DECam images \citep{SVcat},
and the directional light radius method \citep{Sullivan06,Gupta16}.
$\Mhost$ is  derived from 
fitting galaxy model SEDs to $griz$ broadband fluxes with {\tt ZPEG} \citep{ZPEG}.
The SEDs are generated with 
Projet d'Etude des GAlaxies par Synthese Evolutive (\texttt{PEGASE}; \citealt{PEGASE}).
In the DES-SN subset, 116 out of \NDES\ hosts have $\Mhost <10^{10}\Msun$. 
The low-$z$ host galaxy stellar masses are taken from \citet{pantheon}.

\subsection{ $\mu$-Bias Corrections  } 
\label{sect:bias}

We use a simulation-based method (\citealt{Kessler18}) to correct for 
distance biases arising from survey and spectroscopic selection efficiencies,
and also from the analysis and light curve fitting.
For each SN~Ia we calculate the bias correction in Eq.~\ref{eq:mu},
$\dmuBias \equiv \langle\mu - \mutrue \rangle$, 
where $\langle\rangle$ is the average in bins of measured redshift, color, and stretch.
The distance $\mu$ is determined by analyzing the simulated data in the same 
way as the real data (but with $\dmuBias=0$), 
and $\mutrue$ is the true distance modulus used to generate each simulated event.
The correction increases with redshift, and for
individual SNe Ia can be as large as 0.4 mag 
(\S9 of \citealt{Kessler18}).

The simulation accurately models \SAMPLENAME\ selection effects. 
For each generated event it picks a random redshift, color, and stretch
from known distributions \citep{Perrett12,SK16}. Next, it computes true SN~Ia magnitudes at 
all epochs using the SALT2 SED model, intrinsic scatter model (\S\ref{sect:scatter}),
telescope+atmosphere transmission functions for each filter band, 
and cosmological effects such as dimming, redshifting, gravitational lensing,
and galactic extinction.
Using the survey cadence and observing conditions (point spread function, sky noise, zero point), 
instrumental noise is added. 
Finally, our simulation models the efficiencies of \Diff\ and \spec\ confirmation.
The quality of the simulation is illustrated by the good agreement between the 
predicted and observed distribution of many observables including redshift, stretch, and color
(Figs~6 \& 7 in \citealt{Kessler18}, and Fig.~5 in \BSYS).

\subsection{Intrinsic scatter model}
\label{sect:scatter}
We simulate bias corrections with two different models of intrinsic scatter 
that span the range of possibilities in current data samples.
First is the `G10' model, based on \citet{Guy10}, in which
the scatter is primarily achromatic.  Second is the `C11' model, 
based on \citet{Chotard11}, which has stronger scatter in color.
For use in simulations, \citet{Kessler13} converted each of these broadband
scatter models into an SED-variation model.

\subsection{Generating the Bias-Corrected Hubble Diagram}
\label{hubble}

We use the ``BEAMS with Bias Corrections'' (BBC) method \citep{bbc} to 
fit for \{$\alpha,\beta,\gamma,M_0$\}
and to fit for a weighted-average bias-corrected $\mu$ in \NZBIN\ redshift bins.
In addition to propagating the uncertainty from each term in Eq.~\ref{eq:mu},
the BBC fit adds an empirically determined $\mu$-uncertainty ($\sigint$) to each event
so that the best fit $\chi^2/N_{\rm dof}=1$.
This redshift-binned Hubble diagram is used for cosmology fitting as described in \S\ref{subsec:cosmoFit}.
Fig.~\ref{fig:hubble} shows the binned Hubble diagram, and also the
unbinned Hubble diagram using individual bias-corrected distances computed in the BBC fit.

\subsection{Cosmology Fitting}
\label{subsec:cosmoFit}
Cosmological parameters are constrained using the log-likelihood
\begin{equation}
\chi^2 = \vecD^T~[\Ctot]^{-1}~\vecD
\label{eq:cosmochi2}
\end{equation}
and minimizing the posterior with
CosmoMC (\citealt{cosmomc}).
$D_i = \muzDATA - \muzMODEL$ for redshift bin $i=1,\NZBIN$, 
$\muzDATA$ is the BBC-fitted distance modulus in the $i$'th redshift bin, and
$\muzMODEL$ is given by Eq.~\ref{eq:mumodel}.
The covariance matrix ($\Ctot$) is described in \S3.8.2 of \BSYS,
and incorporates systematic uncertainties from each analysis component in \S\ref{sect:ana}.

$\vecD$ and $\Ctot$ are computed separately using the G10 and C11 scatter model 
in the bias-correction simulation.
Each set of quantities is averaged over the G10 and C11 models, 
and these averages are used in Eq.~\ref{eq:cosmochi2}. 
The purpose of averaging is to mitigate the systematic uncertainty related to our 
understanding of intrinsic scatter (\S4.2 of \BSYS). 

Finally, we combine these SN Ia results with priors from CMB and BAO
as described in \S\ref{sect:results}.

\subsection{Blinding and Validation}
The cosmological parameters were blinded until
preliminary results were presented at the 
231st meeting of the American Astronomical Society in January \citeyear{aas231}.
The criteria for unblinding (\S7 of \BSYS)
included analyzing large simulated \SAMPLENAME\ data sets,
and requiring 
i) $w$-bias below 0.01, and 
ii) the rms of $w$-values agrees with the fitted $w$-uncertainty,
for simulations with and without systematic variations. 
Following this initial unblinding, several updates were performed 
(\S3.8.4 of \BSYS), again blinded,
and the final results presented here were unblinded during the DES internal review process.
Compared to the initial unblinding, $w$ increased by \wdifUnblind\ and the 
total uncertainty increased by 3\% (0.057 to 0.059).

\begin{figure}
\begin{centering}
\includegraphics[width=0.47\textwidth]{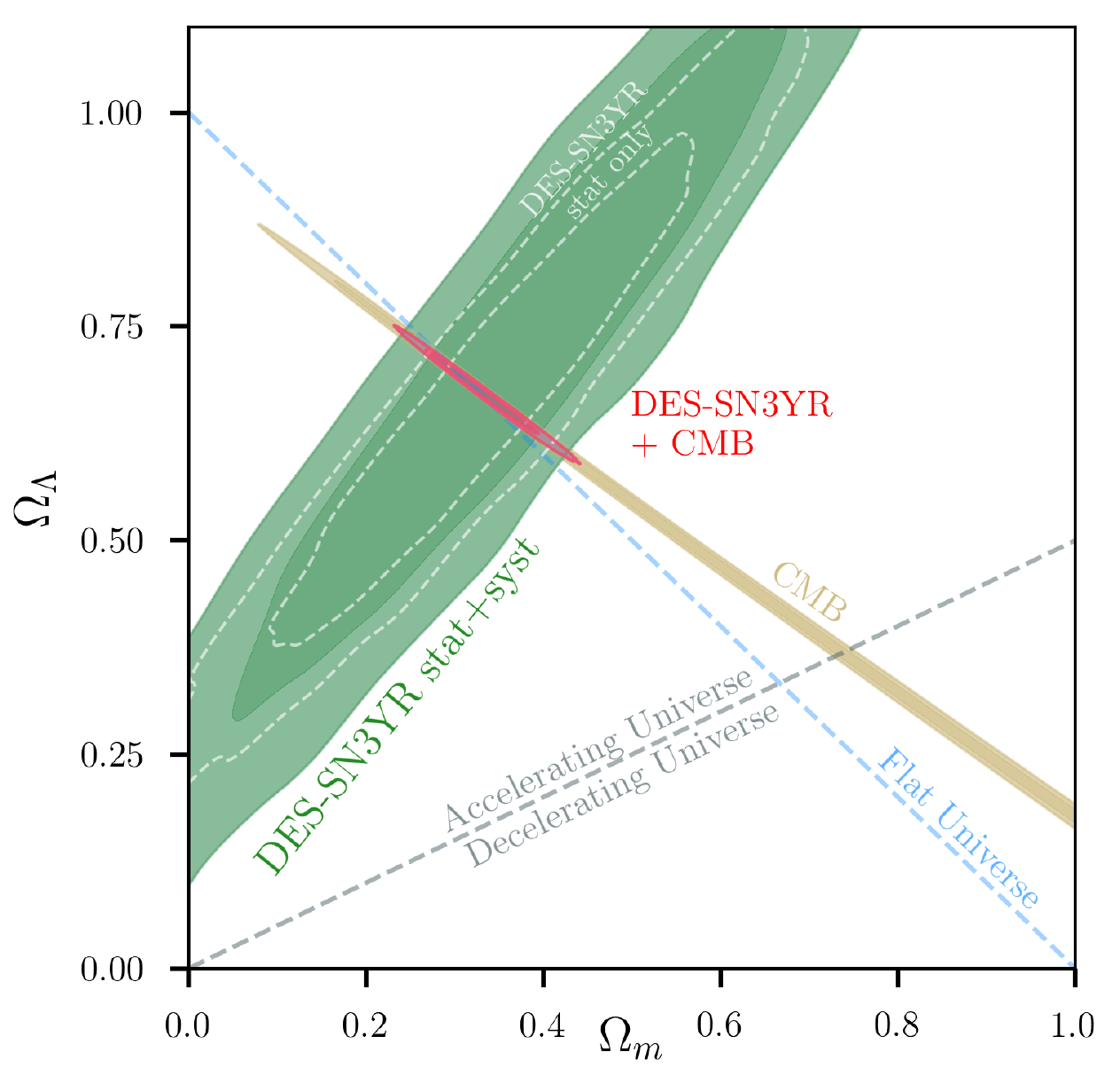}
\caption{Constraints on $\om$-$\oll$ for \LCDM\ model (68\% and 95\% confidence intervals). 
SN contours are shown with statistical uncertainty only (white-dashed),
and with total uncertainty (green shaded). 
Constraints from CMB (brown) and \SAMPLENAME+CMB combined (red), are also shown.\\}
\label{fig:oml}
\end{centering}
\end{figure}

\section{Results}\label{sect:results}

We present the first cosmological results using SNe~Ia from DES. 
We begin with the BBC-fitted parameters ($\alpha,\beta,\gamma,\sigint$) 
in \S\ref{sect:BBCresults}, 
then present our statistical and systematic uncertainty budget for $w$
in \S\ref{sect:werr} and Table~\ref{Tab:finalbudget}.
Finally, we present our cosmological parameters in \S\ref{sect:cosmomodels} and Table~\ref{Tab:wCDM}.  
For our primary results we combine DES-SN3YR with the CMB likelihood
from \citealt{planck16} using their temperature power spectrum and 
low-$\ell$ polarization results.
We also present results without a CMB prior, and with 
both CMB and BAO priors.
All reported uncertainties correspond to 68\% confidence.
To evaluate consistency between our primary result and BAO, 
we compute the evidence using \texttt{PolyChord} 
\citep{Polychord2015a,Polychord2015b},
and compute the evidence ratio ($R$) defined in Eq.~V.3 of \cite{desy1cosmo}. 
Consistency is defined by $R>0.1$.

\subsection{Results for Standardization Parameters}
\label{sect:BBCresults}

While the cosmology results are based on averaging distances using
the G10 and C11 intrinsic scatter models, 
here we show best-fit BBC values from {\BSYS}
using the G10 intrinsic scatter model:
$\alpha = 0.146\pm0.009$, 
$\beta = 3.03 \pm 0.11$, 
$\gamma = 0.025 \pm 0.018$, and 
$\sigint=0.094\pm0.008$.  
Our $\alpha$, $\beta$, and $\sigint$ values
are consistent with  those  found in previous analyses, 
while $\gamma$ is smaller
compared to those in \citet{Kelly10,Sullivan10,Lampeitl10,Betoule14,pantheon}.
Results with the C11 model (Table~5 of \BSYS) show similar trends.

We also check the consistency among the DES-SN and low-$z$ subsets.
While $\alpha$ and $\beta$ are consistent, 
we find $\sigint=0.066\pm0.006$ for DES-SN,
the lowest value of any rolling SN survey.
This value differs by $3.3\sigma$ from
$\sigint=0.120\pm0.015$ for the low-$z$ subset, and the
systematic uncertainty in adopting a single $\sigint$ value 
is discussed below in \S\ref{sect:werr} and also in \S7.3 of \BSYS.
Our $\gamma$ values differ by $1.5\sigma$:
$\gamma_{\rm DES} = 0.009 \pm 0.018$ (consistent with zero) and
$\gamma_{{\rm low}z} = 0.070 \pm 0.038$.

\subsection{$w$ Uncertainty Budget}
\label{sect:werr}

Contributions to the systematic uncertainty budget are presented in \BSYS\ and shown here in Table~\ref{Tab:finalbudget} for flat \WCDM\ fits combined with the CMB likelihood.
The statistical uncertainty on $w$ ($\sigwstat$) is determined without
systematic contributions.
Each systematic contribution is defined as
\begin{equation}
    \sigwsyst = \sqrt{ (\sigwtot)^2 - (\sigwstat)^2 } 
\end{equation}
where $\sigwtot$ is the total (stat+syst) uncertainty 
from including a specific systematic, or a group of systematics.
The uncertainty in $w$ has nearly equal contributions from statistical 
and systematic uncertainties,
the latter of which is broken into four groups in Table~\ref{Tab:finalbudget}.

The first three systematic groups have nearly equal contributions:
1) photometry and calibration ($\sigma_w=\werrSystCal$), 
which includes uncertainties from
the DES-SN and low-$z$ subsets, 
data used to train the SALT2 lightcurve model, and
the HST Calspec standard,
2) $\mu$-bias corrections from the survey ($\sigma_w=\werrSystBiasSurvey$),
which includes uncertainties from rejecting Hubble residual outliers in the low-$z$ subset, 
magnitude versus volume limited selection for low-$z$, 
DES-SN \spec\ selection efficiency, and
determination of DES-SN flux uncertainties, and
3) $\mu$-bias corrections from astrophysical effects ($\sigma_w=\werrSystBiasAstro$),
which includes uncertainties from intrinsic scatter modeling
(G10 vs.\ C11, and two $\sigint$,
parent populations of stretch and color, 
choice of $w$ and $\om$ in the simulation,
and Galactic extinction.
The $4^{th}$ systematics group, redshift ($\sigma_w = \werrSystRedshift$),
includes a global shift in the redshift and peculiar velocity corrections.

Finally, the Table~\ref{Tab:finalbudget} systematics marked with a dagger ($\dagger$)
have not been included in previous analyses,
and the combined uncertainty is $\sigma_w = \werrSystNew$.
Most of this new uncertainty is related to the low-$z$ subset, 
which is almost 40\% of the \SAMPLENAME\ sample. 
For previous analyses with a smaller fraction of low-$z$ events (e.g., Pantheon, JLA)
we do not recommend adding the full $\werrSystNew$ $w$-uncertainty to their results.

\begin{deluxetable}{l|cc}
\tablecolumns{3}
\tablewidth{18pc}
\tablecaption{$w$ Uncertainty Contributions for $w$CDM model\tablenotemark{a} }
\tablehead {
\colhead {Description\tablenotemark{b}} &
\colhead { $\sigma_{w}$}               	&
\colhead {$\sigma_{w}/\sigwstat$}              
}
\startdata
		 Total Stat ($\sigwstat$)	&   \werrStat\	&   1.00 \\
         Total Syst\tablenotemark{c}&	\werrSyst\	&	1.00 \\
         Total Stat+Syst			&	\werrTot\ 	&	1.40 \\
\hline \\
   {\bf [Photometry \& Calibration]} & {\bf [\werrSystCal]}& {\bf [0.50]} \\
     Low-$z$ 			& 0.014	&	0.33 \\
     DES				& 0.010	&	0.24 \\
     SALT2 model 		& 0.009	&	0.21 \\
     HST Calspec   		& 0.007	&	0.17 \\     
         & & \\
    {\bf [$\mu$-Bias Correction: survey]} & {\bf  [\werrSystBiasSurvey]} & {\bf [0.55]} \\
   \tablenotemark{$\dagger$}Low-$z$ 3$\sigma$ Cut	& 0.016	&	0.38 \\
   Low-$z$ Volume Limited				& 0.010	&	0.24 \\
   Spectroscopic Efficiency				& 0.007	&	0.17 \\
   \tablenotemark{$\dagger$}Flux Err Modeling	& 0.001	&	0.02 \\
        & & \\         
   {\bf [$\mu$-Bias Correction: astrophysical]} & {\bf [\werrSystBiasAstro}] & {\bf [0.62]} \\
   Intrinsic Scatter Model (G10 vs. C11)	& 0.014	&	0.33 \\
   \tablenotemark{$\dagger$}Two $\sigma_{\rm int}$	& 0.014	&	0.33 \\
   $\cSalt$, $x_1$ Parent Population						& 0.014	&	0.33 \\
   \tablenotemark{$\dagger$}$w,\om$ in sim. 	& 0.006	&	0.14 \\
   MW Extinction							& 0.005	&	0.12 \\
     & & \\
 {\bf [Redshift]} & {\bf [\werrSystRedshift}] & {\bf [0.29]} \\ 
   Peculiar Velocity		    		& 0.007	& 0.17 \\
   \tablenotemark{$\dagger$}$z+0.00004$	& 0.006	& 0.14 \\
\enddata
\tablenotetext{a}{The sample is \SAMPLENAME\  (DES-SN + low-$z$ sample) plus CMB prior.}
\tablenotetext{b}{Item in {\bf [bold]} is a sub-group and its uncertainty.}
\tablenotetext{c}{The quadrature sum of all systematic uncertainties does not equal $\werrSyst$ because of redshift-dependent correlations when using the full covariance matrix.}
\tablenotetext{$\dagger$}{Uncertainty was not included in previous analyses.}
\label{Tab:finalbudget}
\vspace{.1in}
\end{deluxetable}

\subsection{Cosmology results}\label{sect:cosmomodels}

\subsubsection{{\LCDM}} \label{sect:lcdm}

Using \SAMPLENAME\ and assuming a flat $\Lambda$CDM model,
we find $\om=\omLCDMSNFLAT$.
Assuming a \LCDM\ model  with
curvature ($\ok$) added as a free parameter in Eq.~\ref{eq:Hz} 
(e.g., see Sect~3.1 of \citealt{davis2017})
we find the constraints shown in Fig.~2 and Table~\ref{Tab:wCDM} (row~2).  
Solid contours show our result with both statistical and systematic uncertainties included, 
while dashed contours show the statistical-only uncertainties for comparison. 
Fig.~2 also shows that the CMB data provide strong flatness constraints,
consistent with zero curvature;
the impact of using this CMB prior is shown in row~3.
The impact from adding a BAO prior is shown in row~4,
where the evidence ratio $R=\RLCDM$ shows consistency between the SN+CMB and BAO posteriors.

\subsubsection{Flat $w$CDM} \label{sect:wcdm}
\begin{figure}
\begin{centering}
\includegraphics[width=0.46\textwidth]{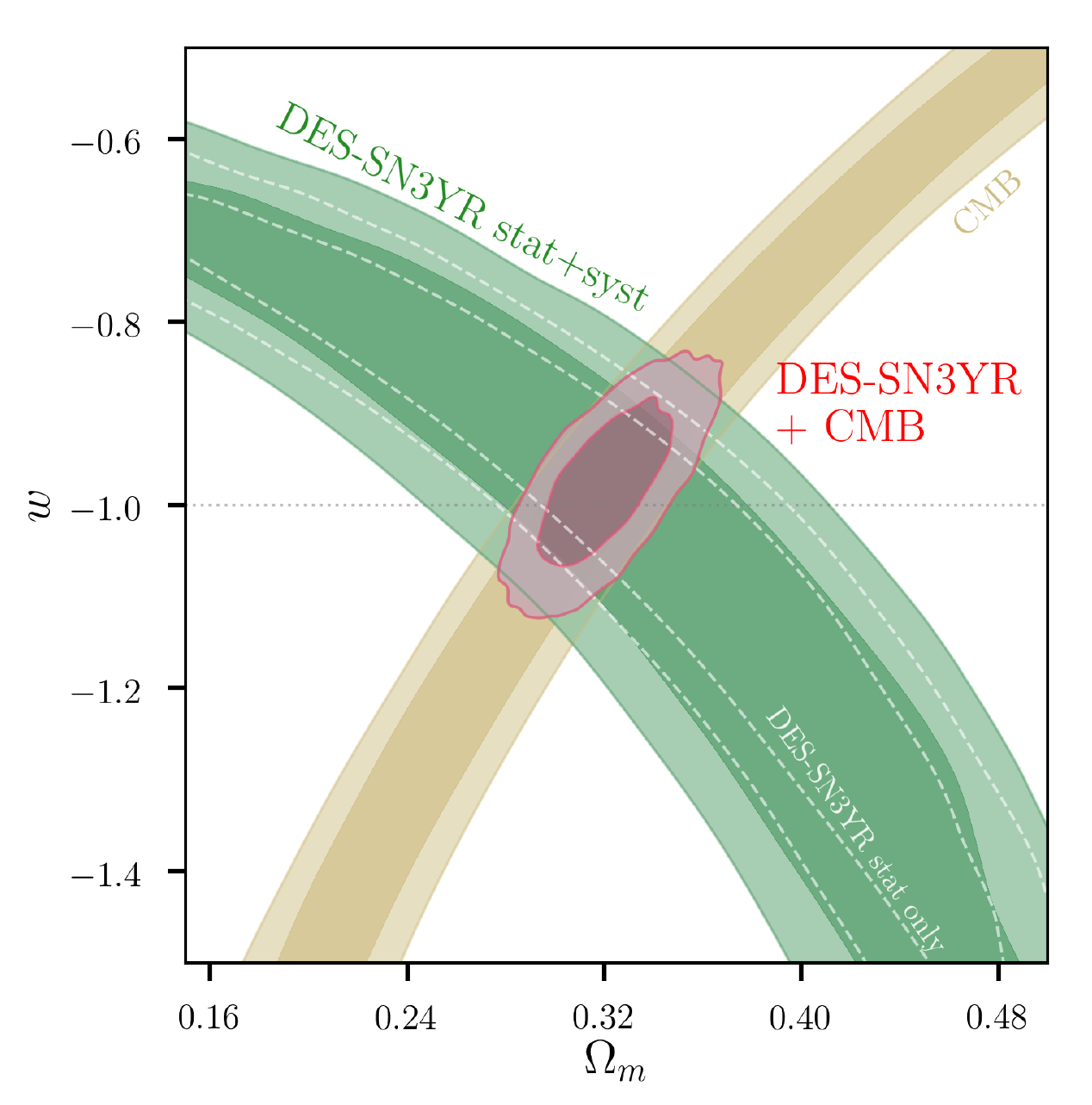}
\caption{Constraints on $\om$-$w$ for the flat $w$CDM model 
(68\% and 95\% confidence intervals).
SN contours are shown with only statistical uncertainty (white-dashed) and 
with total uncertainty (green-shaded). 
Constraints from CMB (brown) and \SAMPLENAME+CMB combined (red) are also shown.\\}
\label{fig:omw}
\end{centering}
\end{figure}

For our primary result, we use \SAMPLENAME\  with the CMB prior and a flat $w$CDM model ($\ok=0$) and find
$\om=\omSNCMB$ and $w=\wSNCMB$ (Table~\ref{Tab:wCDM}, row~5).  Our constraint on $w$ is consistent with the cosmological-constant model for dark energy.  The 68\% and 95\% confidence intervals are given by the red contours in Fig.~\ref{fig:omw}, 
which also shows the contributions from \SAMPLENAME\  and CMB. 
We show two contours for {\SAMPLENAME}, with and without systematic uncertainties in order to demonstrate their impact. In Table~\ref{Tab:wCDM}, row~6, we show the impact of the low-redshift SN sample by removing it; 
the $w$-uncertainty increases by $25\%$ and the constraint lies approximately $1\sigma$ from $w=-1$. 

Next, we consider other combinations of data.  
Adding a BAO prior  (\citealt{BAODR12}; \citealt{MGS}; \citealt{6df}) 
in addition to the CMB prior and SN constraints, 
our best fit $w$-value (Table~2, row~7) is shifted by only $0.006$, the uncertainty is reduced by $\sim20\%$ compared to our primary result, 
and the evidence ratio between SN+CMB and BAO is $R=\RwCDM$ 
showing consistency among the data sets.
If we remove the low-$z$ SN subset (row~8), the $w$-uncertainty increases by only $\sim8\%$. 
Furthermore, we remove the SN sample entirely 
and find that the $w$-uncertainty increases by nearly $50\%$ (row~9).

 \subsubsection{Flat $w_0w_a$CDM}
Our last test is for $w$ evolution using the $w_0w_a$CDM model, 
where $w=w_0 + w_a(1-a)$ and $a=(1+z)^{-1}$. 
Combining probes from SNe, CMB, and BAO,
we find results (Table~2, row~10) consistent with a cosmological constant
($w_0,w_a = -1,0$) and a figure of merit \citep{DETF} of \wowaFOM.
Removing the SN sample increases the $w_0$ and $w_a$ uncertainties by a 
factor of 2 and 1.5, respectively (row~11).

\begin{deluxetable*}{clccc}
\tablecolumns{4}
\tablecaption{Cosmological results\tablenotemark{a}}
\tablehead {
\colhead {Row}            &
\colhead {SN Sample + Prior ({\LCDM}) \phantom{hidanitsme}}            &
\colhead {$\om$}               &
\colhead {$\oll$}  &
\colhead {} 
}
\startdata
1 & \textbf{DES-SN3YR\tablenotemark{b}+flatness} & $\omLCDMSNFLAT$   & $\olLCDMSNFLAT$ \\
2 & DES-SN3YR				           & $\omLCDMSNonly$    & $\olLCDMSNonly$ \\
3 & DES-SN3YR+CMB\tablenotemark{c}     & $\omLCDMSNCMB$    & $\olLCDMSNCMB$  \\
4 & DES-SN3YR+CMB+BAO\tablenotemark{d} & $\omLCDMSNCMBBAO$ & $\olLCDMSNCMBBAO$ \\ 
\\ \hline\hline 
{\white \large{Row}} & SN Sample + Prior (Flat $w$CDM)    & $\om$    & $w$          &  \\  \hline 
5 & \textbf{DES-SN3YR+CMB} {\white\large{R}}          & $\omSNCMB$       & $\wSNCMB$    & \\
6 & DES-SN\tablenotemark{e}+CMB      & $\omDESSNCMB$    & $\wDESSNCMB$    & \\
7 & DES-SN3YR+CMB+BAO                & $\omSNCMBBAO$    & $\wSNCMBBAO$ & \\
8 & DES-SN+CMB+BAO  & $\omDESSNCMBBAO$ & $\wDESSNCMBBAO$ & \\
9 & CMB+BAO                          & $\omCMBBAO$      & $\wCMBBAO$  & \\
\\ \hline\hline 
{\white \large{Row}} & SN Sample + Prior (Flat $w_0w_a$CDM) & $\om$ & $w_0$  & $w_a$ \\  \hline 
10 & \textbf{DES-SN3YR+CMB+BAO}{\white\large{R}} & $\omwowaSNCMBBAO$ & $\woSNCMBBAO$ & $\waSNCMBBAO$ \\
11 & CMB+BAO           & $\omwowaCMBBAO$   & $\woCMBBAO$   & $\waCMBBAO$  \\
\enddata
\tablenotetext{a}{Samples in bold font are primary results given in the abstract.}
\tablenotetext{b}{DES-SN3YR: DES-SN + Low-$z$ samples.}
\tablenotetext{c}{CMB: Planck TT + lowP likelihood (\citealt{planck16}).}
\tablenotetext{d}{BAO: SDSS DR12 (\citealt{BAODR12}); SDSS MGS (\citealt{MGS}); 6dFGS (\citealt{6df})}
\tablenotetext{e}{DES-SN alone (no low-$z$). \\ \\ \\} 
\label{Tab:wCDM}
\end{deluxetable*}

\subsection{Comparison to other SN~Ia Surveys/Analyses}

The \SAMPLENAME\ result has competitive
constraining power given the sample size 
($\sigwtot=\werrTot$ with \NTOT\ total SNe~Ia), 
even after taking into account additional sources of systematic uncertainty.
While our \SAMPLENAME\ sample is $< 1/3$ the size of the Pantheon sample 
(PS1+SNLS+SDSS+low-$z$+HST, $\sigwtot =0.041$), 
our low-$z$ subset is 70\% the size of Pantheon's low-$z$ subset,
and we included five additional sources of systematic uncertainty,
our improvements (\S\ref{sect:intro}) result in a $w$-uncertainty that
is only $\times 1.4$ larger.


\section{Discussion and Conclusion}

We have presented the first cosmological results from the DES-SN program: 
$\om=\omSNCMB$ and $w=\wSNCMB$ for a flat \WCDM\ model after combining with CMB constraints.
These results are consistent with a cosmological constant model and demonstrate the 
high constraining power (per SN) of the DES-SN sample.
\SAMPLENAME\ data products used in this analysis are publicly available at {\urlDR}.
These products include 
filter transmissions, 
redshifts, 
light curves, 
host masses, 
light-curve fit parameters, 
Hubble Diagram,
bias corrections, 
covariance matrix, 
MC chains, 
and code releases.

We have utilized the spectroscopically confirmed SN~Ia sample
from the first three years of DES-SN as well as a low-redshift sample. 
This 3-year sample contains $\sim 10$\%
of the SNe~Ia discovered by DES-SN over the full five year survey.  
Many of the techniques established in this analysis 
will form the basis of upcoming analyses on the much larger
5-year photometrically identified sample.

To benefit from the increased  statistics in the 5-year sample
it will be critical to reduce systematic uncertainties.  
We are working to improve calibration with a large sample of DA White Dwarf observations, including two HST Calspec standards.
Other improvements to systematics are discussed in \S7.2 of \BSYS. 
We are optimistic that our systematic uncertainties can remain at the level of our 
statistical uncertainties for the 5-year analysis. 
This progress in understanding systematics will be critical for making new, 
exciting measurements of dark energy and for paving the way towards 
Stage-IV dark energy experiments like
the Large Synoptic Survey Telescope and the 
Wide Field Infrared Space Telescope.

\acknowledgments

Funding for the DES Projects has been provided by the U.S. Department of Energy, the U.S. National Science Foundation, the Ministry of Science and Education of Spain, 
the Science and Technology Facilities Council of the United Kingdom, the Higher Education Funding Council for England, the National Center for Supercomputing 
Applications at the University of Illinois at Urbana-Champaign, the Kavli Institute of Cosmological Physics at the University of Chicago, 
the Center for Cosmology and Astro-Particle Physics at the Ohio State University,
the Mitchell Institute for Fundamental Physics and Astronomy at Texas A\&M University, Financiadora de Estudos e Projetos, 
Funda{\c c}{\~a}o Carlos Chagas Filho de Amparo {\`a} Pesquisa do Estado do Rio de Janeiro, Conselho Nacional de Desenvolvimento Cient{\'i}fico e Tecnol{\'o}gico and 
the Minist{\'e}rio da Ci{\^e}ncia, Tecnologia e Inova{\c c}{\~a}o, the Deutsche Forschungsgemeinschaft and the Collaborating Institutions in the Dark Energy Survey. 

The Collaborating Institutions are Argonne National Laboratory, the University of California at Santa Cruz, the University of Cambridge, Centro de Investigaciones Energ{\'e}ticas, 
Medioambientales y Tecnol{\'o}gicas-Madrid, the University of Chicago, University College London, the DES-Brazil Consortium, the University of Edinburgh, 
the Eidgen{\"o}ssische Technische Hochschule (ETH) Z{\"u}rich, 
Fermi National Accelerator Laboratory, the University of Illinois at Urbana-Champaign, the Institut de Ci{\`e}ncies de l'Espai (IEEC/CSIC), 
the Institut de F{\'i}sica d'Altes Energies, Lawrence Berkeley National Laboratory, the Ludwig-Maximilians Universit{\"a}t M{\"u}nchen and the associated Excellence Cluster Universe, 
the University of Michigan, the National Optical Astronomy Observatory, the University of Nottingham, The Ohio State University, the University of Pennsylvania, the University of Portsmouth, SLAC National Accelerator Laboratory, Stanford University, the University of Sussex, Texas A\&M University, and the OzDES Membership Consortium.

Based in part on observations at Cerro Tololo Inter-American Observatory, National Optical Astronomy Observatory, which is operated by the Association of Universities for Research in Astronomy (AURA) under a cooperative agreement with the National Science Foundation.

The DES data management system is supported by the National Science Foundation under Grant Numbers AST-1138766 and AST-1536171.
The DES participants from Spanish institutions are partially supported by MINECO under grants AYA2015-71825, ESP2015-66861, FPA2015-68048, SEV-2016-0588, SEV-2016-0597, and MDM-2015-0509, 
some of which include ERDF funds from the European Union. IFAE is partially funded by the CERCA program of the Generalitat de Catalunya.
Research leading to these results has received funding from the European Research
Council under the European Union's Seventh Framework Program (FP7/2007-2013) 
including ERC grant agreements 240672, 291329, 306478, and 615929.
We  acknowledge support from the Australian Research Council Centre of Excellence for All-sky Astrophysics (CAASTRO), through project number CE110001020, and the Brazilian Instituto Nacional de Ci\^enciae Tecnologia (INCT) e-Universe (CNPq grant 465376/2014-2).

This manuscript has been authored by Fermi Research Alliance, LLC under Contract No. DE-AC02-07CH11359 with the U.S. Department of Energy, Office of Science, Office of High Energy Physics. The United States Government retains and the publisher, by accepting the article for publication, acknowledges that the United States Government retains a non-exclusive, paid-up, irrevocable, world-wide license to publish or reproduce the published form of this manuscript, or allow others to do so, for United States Government purposes.

This Letter makes use of observations taken using the Anglo-Australian
Telescope under programs ATAC A/2013B/12 and  NOAO 2013B-0317; the Gemini
Observatory under programs NOAO 2013A-0373/GS-2013B-Q-45, NOAO
2015B-0197/GS-2015B-Q-7, and GS-2015B-Q-8; the Gran Telescopio Canarias
under programs GTC77-13B, GTC70-14B, and GTC101-15B; the Keck Observatory
under programs U063-2013B, U021-2014B, U048-2015B, U038-2016A; the Magellan
Observatory under programs CN2015B-89; the MMT under 2014c-SAO-4,
2015a-SAO-12, 2015c-SAO-21; the South African Large Telescope under
programs 2013-1-RSA\_OTH-023, 2013-2-RSA\_OTH-018, 2014-1-RSA\_OTH-016,
2014-2-SCI-070, 2015-1-SCI-063, and 2015-2-SCI-061; and the Very Large
Telescope under programs ESO 093.A-0749(A), 094.A-0310(B), 095.A-0316(A),
096.A-0536(A), 095.D-0797(A).  

\clearpage

\bibliographystyle{apj}
\bibliography{mybibfile}

\begin{thebibliography}{}
\expandafter\ifx\csname natexlab\endcsname\relax\def\natexlab#1{#1}\fi

\bibitem[{{Abbott} {et~al.}(2018){Abbott}, {Abdalla}, {Allam}, {Amara},
  {Annis}, {Asorey}, {Avila}, {Ballester}, {Banerji}, {Barkhouse}, {Baruah},
  {Baumer}, {Bechtol}, {Becker}, {Benoit-L{\'e}vy}, {Bernstein}, {Bertin},
  {Blazek}, {Bocquet}, {Brooks}, {Brout}, {Buckley-Geer}, {Burke}, {Busti},
  {Campisano}, {Cardiel-Sas}, {Carnero Rosell}, {Carrasco Kind}, {Carretero},
  {Castander}, {Cawthon}, {Chang}, {Chen}, {Conselice}, {Costa}, {Crocce},
  {Cunha}, {D{\textquoteright}Andrea}, {da Costa}, {Das}, {Daues}, {Davis},
  {Davis}, {De Vicente}, {DePoy}, {DeRose}, {Desai}, {Diehl}, {Dietrich},
  {Dodelson}, {Doel}, {Drlica-Wagner}, {Eifler}, {Elliott}, {Evrard}, {Farahi},
  {Fausti Neto}, {Fernand ez}, {Finley}, {Flaugher}, {Foley}, {Fosalba},
  {Friedel}, {Frieman}, {Garc{\'\i}a-Bellido}, {Gaztanaga}, {Gerdes},
  {Giannantonio}, {Gill}, {Glazebrook}, {Goldstein}, {Gower}, {Gruen},
  {Gruendl}, {Gschwend}, {Gupta}, {Gutierrez}, {Hamilton}, {Hartley}, {Hinton},
  {Hislop}, {Hollowood}, {Honscheid}, {Hoyle}, {Huterer}, {Jain}, {James},
  {Jeltema}, {Johnson}, {Johnson}, {Kacprzak}, {Kent}, {Khullar}, {Klein},
  {Kovacs}, {Koziol}, {Krause}, {Kremin}, {Kron}, {Kuehn}, {Kuhlmann},
  {Kuropatkin}, {Lahav}, {Lasker}, {Li}, {Li}, {Liddle}, {Lima}, {Lin},
  {L{\'o}pez-Reyes}, {MacCrann}, {Maia}, {Maloney}, {Manera}, {March},
  {Marriner}, {Marshall}, {Martini}, {McClintock}, {McKay}, {McMahon},
  {Melchior}, {Menanteau}, {Miller}, {Miquel}, {Mohr}, {Morganson}, {Mould},
  {Neilsen}, {Nichol}, {Nogueira}, {Nord}, {Nugent}, {Nunes}, {Ogand o}, {Old},
  {Pace}, {Palmese}, {Paz-Chinch{\'o}n}, {Peiris}, {Percival}, {Petravick},
  {Plazas}, {Poh}, {Pond}, {Porredon}, {Pujol}, {Refregier}, {Reil}, {Ricker},
  {Rollins}, {Romer}, {Roodman}, {Rooney}, {Ross}, {Rykoff}, {Sako}, {Sanchez},
  {Sanchez}, {Santiago}, {Saro}, {Scarpine}, {Scolnic}, {Serrano},
  {Sevilla-Noarbe}, {Sheldon}, {Shipp}, {Silveira}, {Smith}, {Smith}, {Smith},
  {Soares-Santos}, {Sobreira}, {Song}, {Stebbins}, {Suchyta}, {Sullivan},
  {Swanson}, {Tarle}, {Thaler}, {Thomas}, {Thomas}, {Troxel}, {Tucker},
  {Vikram}, {Vivas}, {Walker}, {Wechsler}, {Weller}, {Wester}, {Wolf}, {Wu},
  {Yanny}, {Zenteno}, {Zhang}, {Zuntz}, {DES Collaboration}, {Juneau},
  {Fitzpatrick}, {Nikutta}, {Nidever}, {Olsen}, {Scott}, \& {Data Lab}}]{dr1}
{Abbott}, T.~M.~C., {Abdalla}, F.~B., {Allam}, S., {et~al.} 2018, ApJS, 239, 18

\bibitem[{Abbott {et~al.}(2019)Abbott, Alarcon, Allam, Andersen,
  Andrade-Oliveira, Annis, Asorey, Avila, Bacon, Banik, Bassett, Baxter,
  Bechtol, Becker, Bernstein, Bertin, Blazek, Bridle, Brooks, Brout, Burke,
  Calcino, Camacho, Campos, Carnero~Rosell, Carollo, Carrasco~Kind, Carretero,
  Castander, Cawthon, Challis, Chan, Chang, Childress, Crocce, Cunha, D'Andrea,
  da~Costa, Davis, Davis, De~Vicente, DePoy, DeRose, Desai, Diehl, Dietrich,
  Dodelson, Doel, Drlica-Wagner, Eifler, Elvin-Poole, Estrada, Evrard,
  Fernandez, Flaugher, Foley, Fosalba, Frieman, Galbany, Garc\'{\i}a-Bellido,
  Gatti, Gaztanaga, Gerdes, Giannantonio, Glazebrook, Goldstein, Gruen,
  Gruendl, Gschwend, Gutierrez, Hartley, Hinton, Hollowood, Honscheid,
  Hoormann, Hoyle, Huterer, Jain, James, Jarvis, Jeltema, Kasai, Kent, Kessler,
  Kim, Kokron, Krause, Kron, Kuehn, Kuropatkin, Lahav, Lasker, Lemos, Lewis,
  Li, Lidman, Lima, Lin, Macaulay, MacCrann, Maia, March, Marriner, Marshall,
  Martini, McMahon, Melchior, Menanteau, Miquel, Mohr, Morganson, Muir,
  M\"oller, Neilsen, Nichol, Nord, Ogando, Palmese, Pan, Peiris, Percival,
  Plazas, Porredon, Prat, Romer, Roodman, Rosenfeld, Ross, Rykoff, Samuroff,
  S\'anchez, Sanchez, Scarpine, Schindler, Schubnell, Scolnic, Secco, Serrano,
  Sevilla-Noarbe, Sharp, Sheldon, Smith, Soares-Santos, Sobreira, Sommer,
  Swann, Swanson, Tarle, Thomas, Thomas, Troxel, Tucker, Uddin, Vielzeuf,
  Walker, Wang, Weaverdyck, Wechsler, Weller, Yanny, Zhang, Zhang, \&
  Zuntz}]{desy1cosmo}
Abbott, T. M.~C., Alarcon, A., Allam, S., {et~al.} 2019, Phys. Rev. Lett., 122,
  171301

\bibitem[{{Alam} {et~al.}(2017){Alam}, {Ata}, {Bailey}, {Beutler}, {Bizyaev},
  {Blazek}, {Bolton}, {Brownstein}, {Burden}, {Chuang}, {Comparat}, {Cuesta},
  {Dawson}, {Eisenstein}, {Escoffier}, {Gil-Mar{\'{\i}}n}, {Grieb}, {Hand},
  {Ho}, {Kinemuchi}, {Kirkby}, {Kitaura}, {Malanushenko}, {Malanushenko},
  {Maraston}, {McBride}, {Nichol}, {Olmstead}, {Oravetz}, {Padmanabhan},
  {Palanque-Delabrouille}, {Pan}, {Pellejero-Ibanez}, {Percival}, {Petitjean},
  {Prada}, {Price-Whelan}, {Reid}, {Rodr{\'{\i}}guez-Torres}, {Roe}, {Ross},
  {Ross}, {Rossi}, {Rubi{\~n}o-Mart{\'{\i}}n}, {Saito}, {Salazar-Albornoz},
  {Samushia}, {S{\'a}nchez}, {Satpathy}, {Schlegel}, {Schneider},
  {Sc{\'o}ccola}, {Seo}, {Sheldon}, {Simmons}, {Slosar}, {Strauss}, {Swanson},
  {Thomas}, {Tinker}, {Tojeiro}, {Maga{\~n}a}, {Vazquez}, {Verde}, {Wake},
  {Wang}, {Weinberg}, {White}, {Wood-Vasey}, {Y{\`e}che}, {Zehavi}, {Zhai}, \&
  {Zhao}}]{BAODR12}
{Alam}, S., {Ata}, M., {Bailey}, S., {et~al.} 2017, \mnras, 470, 2617

\bibitem[{{Albrecht} {et~al.}(2006){Albrecht}, {Bernstein}, {Cahn}, {Freedman},
  {Hewitt}, {Hu}, {Huth}, {Kamionkowski}, {Kolb}, {Knox}, {Mather}, {Staggs},
  \& {Suntzeff}}]{DETF}
{Albrecht}, A., {Bernstein}, G., {Cahn}, R., {et~al.} 2006, ArXiv Astrophysics
  e-prints, astro-ph/0609591

\bibitem[{{Astier} {et~al.}(2006){Astier}, {Guy}, {Regnault}, {Pain},
  {Aubourg}, {Balam}, {Basa}, {Carlberg}, {Fabbro}, {Fouchez}, {Hook},
  {Howell}, {Lafoux}, {Neill}, {Palanque-Delabrouille}, {Perrett}, {Pritchet},
  {Rich}, {Sullivan}, {Taillet}, {Aldering}, {Antilogus}, {Arsenijevic},
  {Balland}, {Baumont}, {Bronder}, {Courtois}, {Ellis}, {Filiol}, {Gon{\c
  c}alves}, {Goobar}, {Guide}, {Hardin}, {Lusset}, {Lidman}, {McMahon},
  {Mouchet}, {Mourao}, {Perlmutter}, {Ripoche}, {Tao}, \& {Walton}}]{snls06}
{Astier}, P., {Guy}, J., {Regnault}, N., {et~al.} 2006, \aap, 447, 31

\bibitem[{{Astier} {et~al.}(2013){Astier}, {El Hage}, {Guy}, {Hardin},
  {Betoule}, {Fabbro}, {Fourmanoit}, {Pain}, \& {Regnault}}]{Astier13}
{Astier}, P., {El Hage}, P., {Guy}, J., {et~al.} 2013, \aap, 557, A55

\bibitem[{{Betoule} {et~al.}(2014){Betoule}, {Kessler}, {Guy}, {Mosher},
  {Hardin}, {Biswas}, {Astier}, {El-Hage}, {Konig}, {Kuhlmann}, {Marriner},
  {Pain}, {Regnault}, {Balland}, {Bassett}, {Brown}, {Campbell}, {Carlberg},
  {Cellier-Holzem}, {Cinabro}, {Conley}, {D'Andrea}, {DePoy}, {Doi}, {Ellis},
  {Fabbro}, {Filippenko}, {Foley}, {Frieman}, {Fouchez}, {Galbany}, {Goobar},
  {Gupta}, {Hill}, {Hlozek}, {Hogan}, {Hook}, {Howell}, {Jha}, {Le Guillou},
  {Leloudas}, {Lidman}, {Marshall}, {M{\"o}ller}, {Mour{\~a}o}, {Neveu},
  {Nichol}, {Olmstead}, {Palanque-Delabrouille}, {Perlmutter}, {Prieto},
  {Pritchet}, {Richmond}, {Riess}, {Ruhlmann-Kleider}, {Sako}, {Schahmaneche},
  {Schneider}, {Smith}, {Sollerman}, {Sullivan}, {Walton}, \&
  {Wheeler}}]{Betoule14}
{Betoule}, M., {Kessler}, R., {Guy}, J., {et~al.} 2014, \aap, 568, A22

\bibitem[{{Beutler} {et~al.}(2011){Beutler}, {Blake}, {Colless}, {Jones},
  {Staveley-Smith}, {Campbell}, {Parker}, {Saunders}, \& {Watson}}]{6df}
{Beutler}, F., {Blake}, C., {Colless}, M., {et~al.} 2011, \mnras, 416, 3017

\bibitem[{{Blondin} \& {Tonry}(2007)}]{SNID}
{Blondin}, S., \& {Tonry}, J.~L. 2007, \apj, 666, 1024

\bibitem[{{Bohlin} {et~al.}(2014){Bohlin}, {Gordon}, \& {Tremblay}}]{Bohlin14}
{Bohlin}, R.~C., {Gordon}, K.~D., \& {Tremblay}, P.-E. 2014, \pasp, 126, 711

\bibitem[{{Bonnett} {et~al.}(2016){Bonnett}, {Troxel}, {Hartley}, {Amara},
  {Leistedt}, {Becker}, {Bernstein}, {Bridle}, {Bruderer}, {Busha}, {Carrasco
  Kind}, {Childress}, {Castander}, {Chang}, {Crocce}, {Davis}, {Eifler},
  {Frieman}, {Gangkofner}, {Gaztanaga}, {Glazebrook}, {Gruen}, {Kacprzak},
  {King}, {Kwan}, {Lahav}, {Lewis}, {Lidman}, {Lin}, {MacCrann}, {Miquel},
  {O'Neill}, {Palmese}, {Peiris}, {Refregier}, {Rozo}, {Rykoff}, {Sadeh},
  {S{\'a}nchez}, {Sheldon}, {Uddin}, {Wechsler}, {Zuntz}, {Abbott}, {Abdalla},
  {Allam}, {Armstrong}, {Banerji}, {Bauer}, {Benoit-L{\'e}vy}, {Bertin},
  {Brooks}, {Buckley-Geer}, {Burke}, {Capozzi}, {Carnero Rosell}, {Carretero},
  {Cunha}, {D'Andrea}, {da Costa}, {DePoy}, {Desai}, {Diehl}, {Dietrich},
  {Doel}, {Fausti Neto}, {Fernandez}, {Flaugher}, {Fosalba}, {Gerdes},
  {Gruendl}, {Honscheid}, {Jain}, {James}, {Jarvis}, {Kim}, {Kuehn},
  {Kuropatkin}, {Li}, {Lima}, {Maia}, {March}, {Marshall}, {Martini},
  {Melchior}, {Miller}, {Neilsen}, {Nichol}, {Nord}, {Ogando}, {Plazas},
  {Reil}, {Romer}, {Roodman}, {Sako}, {Sanchez}, {Santiago}, {Smith},
  {Soares-Santos}, {Sobreira}, {Suchyta}, {Swanson}, {Tarle}, {Thaler},
  {Thomas}, {Vikram}, {Walker}, \& {Dark Energy Survey Collaboration}}]{SVcat}
{Bonnett}, C., {Troxel}, M.~A., {Hartley}, W., {et~al.} 2016, \prd, 94, 042005

\bibitem[{{Brout}(2018)}]{aas231}
{Brout}, D. 2018, in American Astronomical Society Meeting Abstracts, Vol. 231,
  American Astronomical Society Meeting Abstracts \#231, 219.04

\bibitem[{{Brout} {et~al.}(2019{\natexlab{a}}){Brout}, {Sako}, {Scolnic},
  {et~al.}}]{Brout18-SMP}
{Brout}, D., {Sako}, M., {Scolnic}, D., {et~al.} 2019{\natexlab{a}}, \apj, 874,
  106

\bibitem[{{Brout} {et~al.}(2019{\natexlab{b}}){Brout}, {Scolnic}, {Kessler},
  {et~al.}}]{Brout18-ANA}
{Brout}, D., {Scolnic}, D., {Kessler}, R., {et~al.} 2019{\natexlab{b}}, \apj,
  874, 150

\bibitem[{{Burke} {et~al.}(2018){Burke}, {Rykoff}, {Allam}, {Annis}, {Bechtol},
  {Bernstein}, {Drlica-Wagner}, {Finley}, {Gruendl}, {James}, {Kent},
  {Kessler}, {Kuhlmann}, {Lasker}, {Li}, {Scolnic}, {Smith}, {Tucker},
  {Wester}, {Yanny}, {Abbott}, {Abdalla}, {Benoit-L{\'e}vy}, {Bertin}, {Carnero
  Rosell}, {Carrasco Kind}, {Carretero}, {Cunha}, {D'Andrea}, {da Costa},
  {Desai}, {Diehl}, {Doel}, {Estrada}, {Garc{\'{\i}}a-Bellido}, {Gruen},
  {Gutierrez}, {Honscheid}, {Kuehn}, {Kuropatkin}, {Maia}, {March}, {Marshall},
  {Melchior}, {Menanteau}, {Miquel}, {Plazas}, {Sako}, {Sanchez}, {Scarpine},
  {Schindler}, {Sevilla-Noarbe}, {Smith}, {Smith}, {Soares-Santos}, {Sobreira},
  {Suchyta}, {Tarle}, {Walker}, \& {DES Collaboration}}]{Burke17}
{Burke}, D.~L., {Rykoff}, E.~S., {Allam}, S., {et~al.} 2018, \aj, 155, 41

\bibitem[{{Carrick} {et~al.}(2015){Carrick}, {Turnbull}, {Lavaux}, \&
  {Hudson}}]{Carrick15}
{Carrick}, J., {Turnbull}, S.~J., {Lavaux}, G., \& {Hudson}, M.~J. 2015,
  \mnras, 450, 317

\bibitem[{{Carter} {et~al.}(2018){Carter}, {Beutler}, {Percival}, {Blake},
  {Koda}, \& {Ross}}]{carter18}
{Carter}, P., {Beutler}, F., {Percival}, W.~J., {et~al.} 2018, \mnras, 481,
  2371

\bibitem[{{Childress} {et~al.}(2017){Childress}, {Lidman}, {Davis}, {Tucker},
  {Asorey}, {Yuan}, {Abbott}, {Abdalla}, {Allam}, {Annis}, {Banerji},
  {Benoit-L{\'e}vy}, {Bernard}, {Bertin}, {Brooks}, {Buckley-Geer}, {Burke},
  {Carnero Rosell}, {Carollo}, {Carrasco Kind}, {Carretero}, {Castander},
  {Cunha}, {da Costa}, {D'Andrea}, {Doel}, {Eifler}, {Evrard}, {Flaugher},
  {Foley}, {Fosalba}, {Frieman}, {Garc{\'{\i}}a-Bellido}, {Glazebrook},
  {Goldstein}, {Gruen}, {Gruendl}, {Gschwend}, {Gupta}, {Gutierrez}, {Hinton},
  {Hoormann}, {James}, {Kessler}, {Kim}, {King}, {Kovacs}, {Kuehn}, {Kuhlmann},
  {Kuropatkin}, {Lagattuta}, {Lewis}, {Li}, {Lima}, {Lin}, {Macaulay}, {Maia},
  {Marriner}, {March}, {Marshall}, {Martini}, {McMahon}, {Menanteau}, {Miquel},
  {Moller}, {Morganson}, {Mould}, {Mudd}, {Muthukrishna}, {Nichol}, {Nord},
  {Ogando}, {Ostrovski}, {Parkinson}, {Plazas}, {Reed}, {Reil}, {Romer},
  {Rykoff}, {Sako}, {Sanchez}, {Scarpine}, {Schindler}, {Schubnell}, {Scolnic},
  {Sevilla-Noarbe}, {Seymour}, {Sharp}, {Smith}, {Soares-Santos}, {Sobreira},
  {Sommer}, {Spinka}, {Suchyta}, {Sullivan}, {Swanson}, {Tarle}, {Uddin},
  {Walker}, {Wester}, \& {Zhang}}]{Childress17}
{Childress}, M.~J., {Lidman}, C., {Davis}, T.~M., {et~al.} 2017, \mnras, 472,
  273

\bibitem[{{Chotard} {et~al.}(2011){Chotard}, {Gangler}, {Aldering},
  {Antilogus}, {Aragon}, {Bailey}, {Baltay}, {Bongard}, {Buton}, {Canto},
  {Childress}, {Copin}, {Fakhouri}, {Hsiao}, {Kerschhaggl}, {Kowalski},
  {Loken}, {Nugent}, {Paech}, {Pain}, {Pecontal}, {Pereira}, {Perlmutter},
  {Rabinowitz}, {Runge}, {Scalzo}, {Smadja}, {Tao}, {Thomas}, {Weaver}, {Wu},
  \& {Nearby Supernova Factory}}]{Chotard11}
{Chotard}, N., {Gangler}, E., {Aldering}, G., {et~al.} 2011, \aap, 529, L4

\bibitem[{{Conley} {et~al.}(2011){Conley}, {Guy}, {Sullivan}, {Regnault},
  {Astier}, {Balland}, {Basa}, {Carlberg}, {Fouchez}, {Hardin}, {Hook},
  {Howell}, {Pain}, {Palanque-Delabrouille}, {Perrett}, {Pritchet}, {Rich},
  {Ruhlmann-Kleider}, {Balam}, {Baumont}, {Ellis}, {Fabbro}, {Fakhouri},
  {Fourmanoit}, {Gonz{\'a}lez-Gait{\'a}n}, {Graham}, {Hudson}, {Hsiao},
  {Kronborg}, {Lidman}, {Mourao}, {Neill}, {Perlmutter}, {Ripoche}, {Suzuki},
  \& {Walker}}]{Conley11}
{Conley}, A., {Guy}, J., {Sullivan}, M., {et~al.} 2011, \apjs, 192, 1

\bibitem[{{Contreras} {et~al.}(2010){Contreras}, {Hamuy}, {Phillips},
  {Folatelli}, {Suntzeff}, {Persson}, {Stritzinger}, {Boldt}, {Gonz{\'a}lez},
  {Krzeminski}, {Morrell}, {Roth}, {Salgado}, {Jos{\'e} Maureira}, {Burns},
  {Freedman}, {Madore}, {Murphy}, {Wyatt}, {Li}, \& {Filippenko}}]{CSP}
{Contreras}, C., {Hamuy}, M., {Phillips}, M.~M., {et~al.} 2010, \aj, 139, 519

\bibitem[{{D'Andrea} {et~al.}(2018){D'Andrea}, {Smith}, {Sullivan},
  {et~al.}}]{DAndrea18}
{D'Andrea}, C., {Smith}, M., {Sullivan}, M., {et~al.} 2018, ArXiv:1811.09565,
  arXiv:1811.09565

\bibitem[{{Davis} \& {Parkinson}(2017)}]{davis2017}
{Davis}, T.~M., \& {Parkinson}, D. 2017, {Characterizing Dark Energy Through
  Supernovae}, ed. A.~W. {Alsabti} \& P.~{Murdin}, 2623

\bibitem[{{Fioc} \& {Rocca-Volmerange}(1997)}]{PEGASE}
{Fioc}, M., \& {Rocca-Volmerange}, B. 1997, \aap, 326, 950

\bibitem[{{Flaugher} {et~al.}(2015){Flaugher}, {Diehl}, {Honscheid}, {Abbott},
  {Alvarez}, {Angstadt}, {Annis}, {Antonik}, {Ballester}, {Beaufore},
  {Bernstein}, {Bernstein}, {Bigelow}, {Bonati}, {Boprie}, {Brooks},
  {Buckley-Geer}, {Campa}, {Cardiel-Sas}, {Castander}, {Castilla}, {Cease},
  {Cela-Ruiz}, {Chappa}, {Chi}, {Cooper}, {da Costa}, {Dede}, {Derylo},
  {DePoy}, {de Vicente}, {Doel}, {Drlica-Wagner}, {Eiting}, {Elliott}, {Emes},
  {Estrada}, {Fausti Neto}, {Finley}, {Flores}, {Frieman}, {Gerdes},
  {Gladders}, {Gregory}, {Gutierrez}, {Hao}, {Holland}, {Holm}, {Huffman},
  {Jackson}, {James}, {Jonas}, {Karcher}, {Karliner}, {Kent}, {Kessler},
  {Kozlovsky}, {Kron}, {Kubik}, {Kuehn}, {Kuhlmann}, {Kuk}, {Lahav}, {Lathrop},
  {Lee}, {Levi}, {Lewis}, {Li}, {Mandrichenko}, {Marshall}, {Martinez},
  {Merritt}, {Miquel}, {Mu{\~n}oz}, {Neilsen}, {Nichol}, {Nord}, {Ogando},
  {Olsen}, {Palaio}, {Patton}, {Peoples}, {Plazas}, {Rauch}, {Reil}, {Rheault},
  {Roe}, {Rogers}, {Roodman}, {Sanchez}, {Scarpine}, {Schindler}, {Schmidt},
  {Schmitt}, {Schubnell}, {Schultz}, {Schurter}, {Scott}, {Serrano}, {Shaw},
  {Smith}, {Soares-Santos}, {Stefanik}, {Stuermer}, {Suchyta}, {Sypniewski},
  {Tarle}, {Thaler}, {Tighe}, {Tran}, {Tucker}, {Walker}, {Wang}, {Watson},
  {Weaverdyck}, {Wester}, {Woods}, {Yanny}, \& {DES Collaboration}}]{decam}
{Flaugher}, B., {Diehl}, H.~T., {Honscheid}, K., {et~al.} 2015, \aj, 150, 150

\bibitem[{{Goldstein} {et~al.}(2015){Goldstein}, {D'Andrea}, {Fischer},
  {Foley}, {Gupta}, {Kessler}, {Kim}, {Nichol}, {Nugent}, {Papadopoulos},
  {Sako}, {Smith}, {Sullivan}, {Thomas}, {Wester}, {Wolf}, {Abdalla},
  {Banerji}, {Benoit-L{\'e}vy}, {Bertin}, {Brooks}, {Carnero Rosell},
  {Castander}, {da Costa}, {Covarrubias}, {DePoy}, {Desai}, {Diehl}, {Doel},
  {Eifler}, {Fausti Neto}, {Finley}, {Flaugher}, {Fosalba}, {Frieman},
  {Gerdes}, {Gruen}, {Gruendl}, {James}, {Kuehn}, {Kuropatkin}, {Lahav}, {Li},
  {Maia}, {Makler}, {March}, {Marshall}, {Martini}, {Merritt}, {Miquel},
  {Nord}, {Ogando}, {Plazas}, {Romer}, {Roodman}, {Sanchez}, {Scarpine},
  {Schubnell}, {Sevilla-Noarbe}, {Smith}, {Soares-Santos}, {Sobreira},
  {Suchyta}, {Swanson}, {Tarle}, {Thaler}, \& {Walker}}]{autoscan}
{Goldstein}, D.~A., {D'Andrea}, C.~B., {Fischer}, J.~A., {et~al.} 2015, \aj,
  150, 82

\bibitem[{{Gupta} {et~al.}(2016){Gupta}, {Kuhlmann}, {Kovacs}, {Spinka},
  {Kessler}, {Goldstein}, {Liotine}, {Pomian}, {D'Andrea}, {Sullivan},
  {Carretero}, {Castander}, {Nichol}, {Finley}, {Fischer}, {Foley}, {Kim},
  {Papadopoulos}, {Sako}, {Scolnic}, {Smith}, {Tucker}, {Uddin}, {Wolf},
  {Yuan}, {Abbott}, {Abdalla}, {Benoit-L{\'e}vy}, {Bertin}, {Brooks}, {Carnero
  Rosell}, {Carrasco Kind}, {Cunha}, {da Costa}, {Desai}, {Doel}, {Eifler},
  {Evrard}, {Flaugher}, {Fosalba}, {Gazta{\~n}aga}, {Gruen}, {Gruendl},
  {James}, {Kuehn}, {Kuropatkin}, {Maia}, {Marshall}, {Miquel}, {Plazas},
  {Romer}, {S{\'a}nchez}, {Schubnell}, {Sevilla-Noarbe}, {Sobreira}, {Suchyta},
  {Swanson}, {Tarle}, {Walker}, \& {Wester}}]{Gupta16}
{Gupta}, R.~R., {Kuhlmann}, S., {Kovacs}, E., {et~al.} 2016, \aj, 152, 154

\bibitem[{{Guy} {et~al.}(2010){Guy}, {Sullivan}, {Conley}, {Regnault},
  {Astier}, {Balland}, {Basa}, {Carlberg}, {Fouchez}, {Hardin}, {Hook},
  {Howell}, {Pain}, {Palanque-Delabrouille}, {Perrett}, {Pritchet}, {Rich},
  {Ruhlmann-Kleider}, {Balam}, {Baumont}, {Ellis}, {Fabbro}, {Fakhouri},
  {Fourmanoit}, {Gonz{\'a}lez-Gait{\'a}n}, {Graham}, {Hsiao}, {Kronborg},
  {Lidman}, {Mourao}, {Perlmutter}, {Ripoche}, {Suzuki}, \& {Walker}}]{Guy10}
{Guy}, J., {Sullivan}, M., {Conley}, A., {et~al.} 2010, \aap, 523, A7

\bibitem[{{Handley} {et~al.}(2015{\natexlab{a}}){Handley}, {Hobson}, \&
  {Lasenby}}]{Polychord2015a}
{Handley}, W.~J., {Hobson}, M.~P., \& {Lasenby}, A.~N. 2015{\natexlab{a}},
  \mnras, 450, L61

\bibitem[{{Handley} {et~al.}(2015{\natexlab{b}}){Handley}, {Hobson}, \&
  {Lasenby}}]{Polychord2015b}
---. 2015{\natexlab{b}}, \mnras, 453, 4384

\bibitem[{{Hicken} {et~al.}(2009{\natexlab{a}}){Hicken}, {Wood-Vasey},
  {Blondin}, {Challis}, {Jha}, {Kelly}, {Rest}, \& {Kirshner}}]{Hicken09b}
{Hicken}, M., {Wood-Vasey}, W.~M., {Blondin}, S., {et~al.} 2009{\natexlab{a}},
  \apj, 700, 1097

\bibitem[{{Hicken} {et~al.}(2009{\natexlab{b}}){Hicken}, {Challis}, {Jha},
  {Kirshner}, {Matheson}, {Modjaz}, {Rest}, {Wood-Vasey}, {Bakos}, {Barton},
  {Berlind}, {Bragg}, {Brice{\~n}o}, {Brown}, {Caldwell}, {Calkins}, {Cho},
  {Ciupik}, {Contreras}, {Dendy}, {Dosaj}, {Durham}, {Eriksen}, {Esquerdo},
  {Everett}, {Falco}, {Fernandez}, {Gaba}, {Garnavich}, {Graves}, {Green},
  {Groner}, {Hergenrother}, {Holman}, {Hradecky}, {Huchra}, {Hutchison},
  {Jerius}, {Jordan}, {Kilgard}, {Krauss}, {Luhman}, {Macri}, {Marrone},
  {McDowell}, {McIntosh}, {McNamara}, {Megeath}, {Mochejska}, {Munoz},
  {Muzerolle}, {Naranjo}, {Narayan}, {Pahre}, {Peters}, {Peterson}, {Rines},
  {Ripman}, {Roussanova}, {Schild}, {Sicilia-Aguilar}, {Sokoloski}, {Smalley},
  {Smith}, {Spahr}, {Stanek}, {Barmby}, {Blondin}, {Stubbs}, {Szentgyorgyi},
  {Torres}, {Vaz}, {Vikhlinin}, {Wang}, {Westover}, {Woods}, \& {Zhao}}]{CfA3}
{Hicken}, M., {Challis}, P., {Jha}, S., {et~al.} 2009{\natexlab{b}}, \apj, 700,
  331

\bibitem[{{Hicken} {et~al.}(2012){Hicken}, {Challis}, {Kirshner}, {Rest},
  {Cramer}, {Wood-Vasey}, {Bakos}, {Berlind}, {Brown}, {Caldwell}, {Calkins},
  {Currie}, {de Kleer}, {Esquerdo}, {Everett}, {Falco}, {Fernandez},
  {Friedman}, {Groner}, {Hartman}, {Holman}, {Hutchins}, {Keys}, {Kipping},
  {Latham}, {Marion}, {Narayan}, {Pahre}, {Pal}, {Peters}, {Perumpilly},
  {Ripman}, {Sipocz}, {Szentgyorgyi}, {Tang}, {Torres}, {Vaz}, {Wolk}, \&
  {Zezas}}]{CfA4}
{Hicken}, M., {Challis}, P., {Kirshner}, R.~P., {et~al.} 2012, \apjs, 200, 12

\bibitem[{{Hinton} {et~al.}(2019){Hinton}, {Kim}, {Davis}, {et~al.}}]{Hinton18}
{Hinton}, S., {Kim}, A., {Davis}, T., {et~al.} 2019, \apj, 876, 15

\bibitem[{{Hinton} {et~al.}(2016){Hinton}, {Davis}, {Lidman}, {Glazebrook}, \&
  {Lewis}}]{marz}
{Hinton}, S.~R., {Davis}, T.~M., {Lidman}, C., {Glazebrook}, K., \& {Lewis},
  G.~F. 2016, Astronomy and Computing, 15, 61

\bibitem[{{Holtzman} {et~al.}(2008){Holtzman}, {Marriner}, {Kessler}, {Sako},
  {Dilday}, {Frieman}, {Schneider}, {Bassett}, {Becker}, {Cinabro}, {DeJongh},
  {Depoy}, {Doi}, {Garnavich}, {Hogan}, {Jha}, {Konishi}, {Lampeitl},
  {Marshall}, {McGinnis}, {Miknaitis}, {Nichol}, {Prieto}, {Riess}, {Richmond},
  {Romani}, {Smith}, {Takanashi}, {Tokita}, {van der Heyden}, {Yasuda}, \&
  {Zheng}}]{Holtzman08}
{Holtzman}, J.~A., {Marriner}, J., {Kessler}, R., {et~al.} 2008, \aj, 136, 2306

\bibitem[{{Howell} {et~al.}(2005){Howell}, {Sullivan}, {Perrett}, {Bronder},
  {Hook}, {Astier}, {Aubourg}, {Balam}, {Basa}, {Carlberg}, {Fabbro},
  {Fouchez}, {Guy}, {Lafoux}, {Neill}, {Pain}, {Palanque-Delabrouille},
  {Pritchet}, {Regnault}, {Rich}, {Taillet}, {Knop}, {McMahon}, {Perlmutter},
  \& {Walton}}]{Superfit}
{Howell}, D.~A., {Sullivan}, M., {Perrett}, K., {et~al.} 2005, \apj, 634, 1190

\bibitem[{{Jha} {et~al.}(2006){Jha}, {Kirshner}, {Challis}, {Garnavich},
  {Matheson}, {Soderberg}, {Graves}, {Hicken}, {Alves}, {Arce}, {Balog},
  {Barmby}, {Barton}, {Berlind}, {Bragg}, {Brice{\~n}o}, {Brown}, {Buckley},
  {Caldwell}, {Calkins}, {Carter}, {Concannon}, {Donnelly}, {Eriksen},
  {Fabricant}, {Falco}, {Fiore}, {Garcia}, {G{\'o}mez}, {Grogin}, {Groner},
  {Groot}, {Haisch}, {Hartmann}, {Hergenrother}, {Holman}, {Huchra},
  {Jayawardhana}, {Jerius}, {Kannappan}, {Kim}, {Kleyna}, {Kochanek},
  {Koranyi}, {Krockenberger}, {Lada}, {Luhman}, {Luu}, {Macri}, {Mader},
  {Mahdavi}, {Marengo}, {Marsden}, {McLeod}, {McNamara}, {Megeath}, {Moraru},
  {Mossman}, {Muench}, {Mu{\~n}oz}, {Muzerolle}, {Naranjo}, {Nelson-Patel},
  {Pahre}, {Patten}, {Peters}, {Peters}, {Raymond}, {Rines}, {Schild},
  {Sobczak}, {Spahr}, {Stauffer}, {Stefanik}, {Szentgyorgyi}, {Tollestrup},
  {V{\"a}is{\"a}nen}, {Vikhlinin}, {Wang}, {Willner}, {Wolk}, {Zajac}, {Zhao},
  \& {Stanek}}]{CFAjha}
{Jha}, S., {Kirshner}, R.~P., {Challis}, P., {et~al.} 2006, \aj, 131, 527

\bibitem[{{Kelly} {et~al.}(2010){Kelly}, {Hicken}, {Burke}, {Mandel}, \&
  {Kirshner}}]{Kelly10}
{Kelly}, P.~L., {Hicken}, M., {Burke}, D.~L., {Mandel}, K.~S., \& {Kirshner},
  R.~P. 2010, \apj, 715, 743

\bibitem[{{Kessler} {et~al.}(2019){Kessler}, {Brout}, {Crawford},
  {et~al.}}]{Kessler18}
{Kessler}, R., {Brout}, D., {Crawford}, S., {et~al.} 2019, \mnras, 485, 1171

\bibitem[{{Kessler} \& {Scolnic}(2017)}]{bbc}
{Kessler}, R., \& {Scolnic}, D. 2017, \apj, 836, 56

\bibitem[{{Kessler} {et~al.}(2009){Kessler}, {Bernstein}, {Cinabro}, {Dilday},
  {Frieman}, {Jha}, {Kuhlmann}, {Miknaitis}, {Sako}, {Taylor}, \&
  {Vanderplas}}]{Kessler09SNANA}
{Kessler}, R., {Bernstein}, J.~P., {Cinabro}, D., {et~al.} 2009, \pasp, 121,
  1028

\bibitem[{{Kessler} {et~al.}(2013){Kessler}, {Guy}, {Marriner}, {Betoule},
  {Brinkmann}, {Cinabro}, {El-Hage}, {Frieman}, {Jha}, {Mosher}, \&
  {Schneider}}]{Kessler13}
{Kessler}, R., {Guy}, J., {Marriner}, J., {et~al.} 2013, \apj, 764, 48

\bibitem[{{Kessler} {et~al.}(2015){Kessler}, {Marriner}, {Childress},
  {Covarrubias}, {D'Andrea}, {Finley}, {Fischer}, {Foley}, {Goldstein},
  {Gupta}, {Kuehn}, {Marcha}, {Nichol}, {Papadopoulos}, {Sako}, {Scolnic},
  {Smith}, {Sullivan}, {Wester}, {Yuan}, {Abbott}, {Abdalla}, {Allam},
  {Benoit-L{\'e}vy}, {Bernstein}, {Bertin}, {Brooks}, {Carnero Rosell},
  {Carrasco Kind}, {Castander}, {Crocce}, {da Costa}, {Desai}, {Diehl},
  {Eifler}, {Fausti Neto}, {Flaugher}, {Frieman}, {Gerdes}, {Gruen}, {Gruendl},
  {Honscheid}, {James}, {Kuropatkin}, {Li}, {Maia}, {Marshall}, {Martini},
  {Miller}, {Miquel}, {Nord}, {Ogando}, {Plazas}, {Reil}, {Romer}, {Roodman},
  {Sanchez}, {Sevilla-Noarbe}, {Smith}, {Soares-Santos}, {Sobreira}, {Tarle},
  {Thaler}, {Thomas}, {Tucker}, {Walker}, \& {DES Collaboration}}]{Kessler15}
{Kessler}, R., {Marriner}, J., {Childress}, M., {et~al.} 2015, \aj, 150, 172

\bibitem[{{Lampeitl} {et~al.}(2010){Lampeitl}, {Smith}, {Nichol}, {Bassett},
  {Cinabro}, {Dilday}, {Foley}, {Frieman}, {Garnavich}, {Goobar}, {Im}, {Jha},
  {Marriner}, {Miquel}, {Nordin}, {{\"O}stman}, {Riess}, {Sako}, {Schneider},
  {Sollerman}, \& {Stritzinger}}]{Lampeitl10}
{Lampeitl}, H., {Smith}, M., {Nichol}, R.~C., {et~al.} 2010, \apj, 722, 566

\bibitem[{{Lasker} {et~al.}(2019){Lasker}, {Kessler}, {Scolnic},
  {et~al.}}]{Lasker18}
{Lasker}, J., {Kessler}, R., {Scolnic}, D., {et~al.} 2019, \mnras, 485, 5329

\bibitem[{{Le Borgne} \& {Rocca-Volmerange}(2002)}]{ZPEG}
{Le Borgne}, D., \& {Rocca-Volmerange}, B. 2002, \aap, 386, 446

\bibitem[{Lewis \& Bridle(2002)}]{cosmomc}
Lewis, A., \& Bridle, S. 2002, Phys. Rev., D66, 103511

\bibitem[{{Macaulay} {et~al.}(2019){Macaulay}, {Nichol}, {Bacon},
  {et~al.}}]{Macaulay18}
{Macaulay}, E., {Nichol}, R.~C., {Bacon}, D., {et~al.} 2019, \mnras, 966

\bibitem[{{Morganson} {et~al.}(2018){Morganson}, {Gruendl}, {Menanteau},
  {Carrasco Kind}, {Chen}, {Daues}, {Drlica-Wagner}, {Friedel}, {Gower},
  {Johnson}, {Johnson}, {Kessler}, {Paz-Chinch{\'o}n}, {Petravick}, {Pond},
  {Yanny}, {Allam}, {Armstrong}, {Barkhouse}, {Bechtol}, {Benoit-L{\'e}vy},
  {Bernstein}, {Bertin}, {Buckley-Geer}, {Covarrubias}, {Desai}, {Diehl},
  {Goldstein}, {Gruen}, {Li}, {Lin}, {Marriner}, {Mohr}, {Neilsen}, {Ngeow},
  {Paech}, {Rykoff}, {Sako}, {Sevilla-Noarbe}, {Sheldon}, {Sobreira}, {Tucker},
  {Wester}, \& {DES Collaboration}}]{Morganson18}
{Morganson}, E., {Gruendl}, R.~A., {Menanteau}, F., {et~al.} 2018, \pasp, 130,
  074501

\bibitem[{{Oke} \& {Gunn}(1983)}]{oke}
{Oke}, J.~B., \& {Gunn}, J.~E. 1983, \apj, 266, 713

\bibitem[{{Perlmutter} {et~al.}(1999){Perlmutter}, {Aldering}, {Goldhaber},
  {Knop}, {Nugent}, {Castro}, {Deustua}, {Fabbro}, {Goobar}, {Groom}, {Hook},
  {Kim}, {Kim}, {Lee}, {Nunes}, {Pain}, {Pennypacker}, {Quimby}, {Lidman},
  {Ellis}, {Irwin}, {McMahon}, {Ruiz-Lapuente}, {Walton}, {Schaefer}, {Boyle},
  {Filippenko}, {Matheson}, {Fruchter}, {Panagia}, {Newberg}, {Couch}, \&
  {Project}}]{Perlmutter99}
{Perlmutter}, S., {Aldering}, G., {Goldhaber}, G., {et~al.} 1999, \apj, 517,
  565

\bibitem[{{Perrett} {et~al.}(2012){Perrett}, {Sullivan}, {Conley},
  {et~al.}}]{Perrett12}
{Perrett}, K., {Sullivan}, M., {Conley}, A., {et~al.} 2012, \aj, 144, 59

\bibitem[{{Planck Collaboration} {et~al.}(2016){Planck Collaboration}, {Ade},
  {Aghanim}, {Arnaud}, {Ashdown}, {Aumont}, {Baccigalupi}, {Banday},
  {Barreiro}, {Bartlett}, \& et~al.}]{planck16}
{Planck Collaboration}, {Ade}, P.~A.~R., {Aghanim}, N., {et~al.} 2016, \aap,
  594, A13

\bibitem[{{Rest} {et~al.}(2014){Rest}, {Scolnic}, {Foley}, {Huber}, {Chornock},
  {Narayan}, {Tonry}, {Berger}, {Soderberg}, {Stubbs}, {Riess}, {Kirshner},
  {Smartt}, {Schlafly}, {Rodney}, {Botticella}, {Brout}, {Challis}, {Czekala},
  {Drout}, {Hudson}, {Kotak}, {Leibler}, {Lunnan}, {Marion}, {McCrum},
  {Milisavljevic}, {Pastorello}, {Sanders}, {Smith}, {Stafford}, {Thilker},
  {Valenti}, {Wood-Vasey}, {Zheng}, {Burgett}, {Chambers}, {Denneau}, {Draper},
  {Flewelling}, {Hodapp}, {Kaiser}, {Kudritzki}, {Magnier}, {Metcalfe},
  {Price}, {Sweeney}, {Wainscoat}, \& {Waters}}]{Rest14}
{Rest}, A., {Scolnic}, D., {Foley}, R.~J., {et~al.} 2014, \apj, 795, 44

\bibitem[{{Riess} {et~al.}(1998){Riess}, {Filippenko}, {Challis},
  {Clocchiatti}, {Diercks}, {Garnavich}, {Gilliland}, {Hogan}, {Jha},
  {Kirshner}, {Leibundgut}, {Phillips}, {Reiss}, {Schmidt}, {Schommer},
  {Smith}, {Spyromilio}, {Stubbs}, {Suntzeff}, \& {Tonry}}]{Riess98}
{Riess}, A.~G., {Filippenko}, A.~V., {Challis}, P., {et~al.} 1998, \aj, 116,
  1009

\bibitem[{{Ross} {et~al.}(2015){Ross}, {Samushia}, {Howlett}, {Percival},
  {Burden}, \& {Manera}}]{MGS}
{Ross}, A.~J., {Samushia}, L., {Howlett}, C., {et~al.} 2015, \mnras, 449, 835

\bibitem[{{Scolnic} \& {Kessler}(2016)}]{SK16}
{Scolnic}, D., \& {Kessler}, R. 2016, \apjl, 822, L35

\bibitem[{{Scolnic} {et~al.}(2015){Scolnic}, {Casertano}, {Riess}, {Rest},
  {Schlafly}, {Foley}, {Finkbeiner}, {Tang}, {Burgett}, {Chambers}, {Draper},
  {Flewelling}, {Hodapp}, {Huber}, {Kaiser}, {Kudritzki}, {Magnier},
  {Metcalfe}, \& {Stubbs}}]{scolnic15}
{Scolnic}, D., {Casertano}, S., {Riess}, A., {et~al.} 2015, \apj, 815, 117

\bibitem[{{Scolnic} {et~al.}(2018){Scolnic}, {Jones}, {Rest}, {Pan},
  {Chornock}, {Foley}, {Huber}, {Kessler}, {Narayan}, {Riess}, {Rodney},
  {Berger}, {Brout}, {Challis}, {Drout}, {Finkbeiner}, {Lunnan}, {Kirshner},
  {Sanders}, {Schlafly}, {Smartt}, {Stubbs}, {Tonry}, {Wood-Vasey}, {Foley},
  {Hand}, {Johnson}, {Burgett}, {Chambers}, {Draper}, {Hodapp}, {Kaiser},
  {Kudritzki}, {Magnier}, {Metcalfe}, {Bresolin}, {Gall}, {Kotak}, {McCrum}, \&
  {Smith}}]{pantheon}
{Scolnic}, D.~M., {Jones}, D.~O., {Rest}, A., {et~al.} 2018, \apj, 859, 101

\bibitem[{{Stritzinger} {et~al.}(2011){Stritzinger}, {Phillips}, {Boldt},
  {Burns}, {Campillay}, {Contreras}, {Gonzalez}, {Folatelli}, {Morrell},
  {Krzeminski}, {Roth}, {Salgado}, {DePoy}, {Hamuy}, {Freedman}, {Madore},
  {Marshall}, {Persson}, {Rheault}, {Suntzeff}, {Villanueva}, {Li}, \&
  {Filippenko}}]{CSP2}
{Stritzinger}, M.~D., {Phillips}, M.~M., {Boldt}, L.~N., {et~al.} 2011, \aj,
  142, 156

\bibitem[{{Sullivan} {et~al.}(2006){Sullivan}, {Le Borgne}, {Pritchet},
  {Hodsman}, {Neill}, {Howell}, {Carlberg}, {Astier}, {Aubourg}, {Balam},
  {Basa}, {Conley}, {Fabbro}, {Fouchez}, {Guy}, {Hook}, {Pain},
  {Palanque-Delabrouille}, {Perrett}, {Regnault}, {Rich}, {Taillet}, {Baumont},
  {Bronder}, {Ellis}, {Filiol}, {Lusset}, {Perlmutter}, {Ripoche}, \&
  {Tao}}]{Sullivan06}
{Sullivan}, M., {Le Borgne}, D., {Pritchet}, C.~J., {et~al.} 2006, \apj, 648,
  868

\bibitem[{{Sullivan} {et~al.}(2010){Sullivan}, {Conley}, {Howell}, {Neill},
  {Astier}, {Balland}, {Basa}, {Carlberg}, {Fouchez}, {Guy}, {Hardin}, {Hook},
  {Pain}, {Palanque-Delabrouille}, {Perrett}, {Pritchet}, {Regnault}, {Rich},
  {Ruhlmann-Kleider}, {Baumont}, {Hsiao}, {Kronborg}, {Lidman}, {Perlmutter},
  \& {Walker}}]{Sullivan10}
{Sullivan}, M., {Conley}, A., {Howell}, D.~A., {et~al.} 2010, \mnras, 406, 782

\bibitem[{{Wood-Vasey} {et~al.}(2007){Wood-Vasey}, {Miknaitis}, {Stubbs},
  {Jha}, {Riess}, {Garnavich}, {Kirshner}, {Aguilera}, {Becker}, {Blackman},
  {Blondin}, {Challis}, {Clocchiatti}, {Conley}, {Covarrubias}, {Davis},
  {Filippenko}, {Foley}, {Garg}, {Hicken}, {Krisciunas}, {Leibundgut}, {Li},
  {Matheson}, {Miceli}, {Narayan}, {Pignata}, {Prieto}, {Rest}, {Salvo},
  {Schmidt}, {Smith}, {Sollerman}, {Spyromilio}, {Tonry}, {Suntzeff}, \&
  {Zenteno}}]{essence2}
{Wood-Vasey}, W.~M., {Miknaitis}, G., {Stubbs}, C.~W., {et~al.} 2007, \apj,
  666, 694

\bibitem[{{Yuan} {et~al.}(2015){Yuan}, {Lidman}, {Davis}, {Childress},
  {Abdalla}, {Banerji}, {Buckley-Geer}, {Carnero Rosell}, {Carollo},
  {Castander}, {D'Andrea}, {Diehl}, {Cunha}, {Foley}, {Frieman}, {Glazebrook},
  {Gschwend}, {Hinton}, {Jouvel}, {Kessler}, {Kim}, {King}, {Kuehn},
  {Kuhlmann}, {Lewis}, {Lin}, {Martini}, {McMahon}, {Mould}, {Nichol},
  {Norris}, {O'Neill}, {Ostrovski}, {Papadopoulos}, {Parkinson}, {Reed},
  {Romer}, {Rooney}, {Rozo}, {Rykoff}, {Sako}, {Scalzo}, {Schmidt}, {Scolnic},
  {Seymour}, {Sharp}, {Sobreira}, {Sullivan}, {Thomas}, {Tucker}, {Uddin},
  {Wechsler}, {Wester}, {Wilcox}, {Zhang}, {Abbott}, {Allam}, {Bauer},
  {Benoit-L{\'e}vy}, {Bertin}, {Brooks}, {Burke}, {Carrasco Kind},
  {Covarrubias}, {Crocce}, {da Costa}, {DePoy}, {Desai}, {Doel}, {Eifler},
  {Evrard}, {Fausti Neto}, {Flaugher}, {Fosalba}, {Gaztanaga}, {Gerdes},
  {Gruen}, {Gruendl}, {Honscheid}, {James}, {Kuropatkin}, {Lahav}, {Li},
  {Maia}, {Makler}, {Marshall}, {Miller}, {Miquel}, {Ogando}, {Plazas},
  {Roodman}, {Sanchez}, {Scarpine}, {Schubnell}, {Sevilla-Noarbe}, {Smith},
  {Soares-Santos}, {Suchyta}, {Swanson}, {Tarle}, {Thaler}, \&
  {Walker}}]{Yuan15}
{Yuan}, F., {Lidman}, C., {Davis}, T.~M., {et~al.} 2015, \mnras, 452, 3047

\end{thebibliography}

\clearpage

\end{document}